\documentclass[11pt]{article}
\usepackage{draft}
\usepackage{cite}
\usepackage{cancel}
\usepackage{mathrsfs}
\usepackage{enumitem}
\usepackage{xcolor}
\usepackage{caption}  
\usepackage{graphicx} 
\usepackage{float} 
\usepackage{cite}
\usepackage{relsize}
\usepackage{physics}
\usepackage{psfrag}
\usepackage{cancel}
\usepackage{array}
\usepackage{amssymb}
\usepackage{amsmath}
\usepackage{mathrsfs}
\usepackage[compat=1.1.0]{tikz-feynman}
\usepackage{amsthm}
\usepackage{float}
\usepackage{tikz}
\usepackage{bm}
\usetikzlibrary{patterns}
\usetikzlibrary{mindmap,decorations.pathmorphing,backgrounds,positioning,fit}
\usetikzlibrary{decorations.markings}
\usepackage{tikz,lipsum,lmodern}
\usepackage[most]{tcolorbox}
\usepackage{hyperref}
\usepackage{xcolor}
\usepackage{tikz}
\usetikzlibrary{decorations.pathmorphing,patterns}
\usepackage{amsmath}
\usepackage{amssymb}
\usepackage{float}
\usepackage{fixmath}
\usepackage{physics}
\usepackage{slashed}
\usepackage{graphicx}
\usepackage{mathrsfs}
\usepackage{amsbsy}
\usepackage{subfig}
\usepackage{multirow}
\usepackage{hyperref}
\usepackage{bbm}
\usepackage{tikz}
\usetikzlibrary{arrows.meta,calc}
\definecolor{LightGray}{rgb}{0.8, 0.8, 0.8}

\definecolor{twilightlavender}{rgb}{0.54, 0.29, 0.42}
\definecolor{richmaroon}{rgb}{0.69, 0.19, 0.38}
\definecolor{forestgreen(web)}{rgb}{0.13, 0.55, 0.13}
\definecolor{lava}{rgb}{0.81, 0.06, 0.13}
\hypersetup{
	breaklinks,
	colorlinks,
	citecolor=forestgreen(web),
	filecolor=richmaroon,
	linkcolor=lava,
	urlcolor=twilightlavender
}

\usepackage{cleveref}

\crefformat{section}{\S#2#1#3} 
\crefformat{subsection}{\S#2#1#3}
\crefformat{subsubsection}{\S#2#1#3}

\usepackage{verbatim}

\usepackage{color}

\title{Li\'enard--Wiechert fields in AdS and flat-space antipodal matching from geodesic-centered Coulombic data}

\affiliation[]{Yau Mathematical Sciences Center (YMSC), Tsinghua University, Beijing 100084, China}
\usepackage{orcidlink}
\author[\orcidlink{0000-0002-4535-3198}]{Sarthak Duary}\emailAdd{sarthakduary@tsinghua.edu.cn}

\abstract{We present a geometric derivation of Li\'enard--Wiechert fields in flat-space and AdS, emphasizing the origin of antipodal matching. In flat-space, the field of a uniformly moving charge is rewritten in coordinates centered on the source timelike geodesic. In this frame the charge is at rest and the solution is Coulombic, so the matching of the leading data at null infinity arises from describing a static field in a non-centered frame. We extend this construction to global AdS, where uniform motion is replaced by motion along a timelike geodesic. Starting from the static Coulomb solution at the center, we reconstruct the field of a freely moving charge in arbitrary global coordinates using embedding-space invariants. The resulting closed-form field obeys exact antipodal covariance in the bulk, and its boundary null-fringe limit reproduces the usual flat-space antipodal matching relation. We also describe an image-charge interpretation: the flat-space Coulomb field is represented after conformal compactification by an image singularity at spatial infinity, while the AdS Coulomb seed may be viewed as a charge together with an opposite image charge in a reflected copy. Together, these perspectives give a unified picture of Coulombic Li\'enard--Wiechert fields, antipodal matching, and the AdS-to-flat-space limit.}

\begin{document}
\maketitle

\section{Introduction}

Antipodal matching is the natural gluing condition for long-range fields in asymptotically flat spacetime. In classical electrodynamics, the Li\'enard--Wiechert field of a uniformly moving charge contains a Coulombic component which falls as $1/r^2$. On future null infinity $\mathscr I^+$, with retarded time $u=t-r$, the relevant corner is $\mathscr I^+_-$, the past endpoint of $\mathscr I^+$ corresponding to $u\to -\infty$. Physically, this is the region of future null infinity closest to spatial infinity. Similarly, on past null infinity $\mathscr I^-$, with advanced time $v=t+r$, the relevant corner is $\mathscr I^-_+$, the future endpoint of $\mathscr I^-$ corresponding to $v\to +\infty$. This is the part of past null infinity which also meets spatial infinity. Thus $\mathscr I^+_-$ and $\mathscr I^-_+$ are the two null-infinity corners adjacent to spatial infinity $i^0$. The Coulombic data are therefore
\begin{equation}
 \lim_{r\to\infty} r^2F_{ru}(u,\hat x)\big|_{\mathscr I^+_-},
 \qquad
 \lim_{r\to\infty} r^2F_{rv}(v,\hat x)\big|_{\mathscr I^-_+}.
\end{equation}
These two quantities should not be compared at the same angular point $\hat x$. A null direction which enters spatial infinity from $\mathscr I^-_+$ emerges toward $\mathscr I^+_-$ at the opposite point on the celestial sphere. Therefore the Lorentz-invariant matching through spatial infinity identifies antipodal directions, giving
\begin{equation}
 \lim_{r\to\infty} r^2F_{ru}(u,\hat x)\big|_{\mathscr I^+_-}
 =
 \lim_{r\to\infty} r^2F_{rv}(v,-\hat x)\big|_{\mathscr I^-_+}.
\end{equation}
This condition expresses the continuity of the Coulombic electric flux through spatial infinity. It is the classical gluing condition underlying large-gauge charge conservation, soft photon theorems, electromagnetic memory, and the infrared structure of the flat-space $\mathcal S$-matrix. The Li\'enard--Wiechert field gives the cleanest example of this mechanism: in the rest frame of the charge the field is just Coulombic, but when viewed from a non-centered inertial frame its leading data at $\mathscr I^+_-$ and $\mathscr I^-_+$ agree only after the antipodal identification.

The flat-space limit of AdS/CFT provides a controlled setting in which bulk scattering observables may be extracted from boundary data in the large-radius limit. In particular, \cite{Hijano:2020szl} showed that Li\'enard--Wiechert fields in
AdS can be obtained by starting from the Coulomb field of a static charge and acting with AdS isometries. The Li\'enard--Wiechert potentials give the electromagnetic potentials generated by a point charge moving on an arbitrary worldline. Finding the fields sourced by such a charge is a standard problem in classical electrodynamics. In the usual textbook treatment, one first solves the inhomogeneous wave equation for the retarded potentials and then differentiates these potentials with respect to the observation point and time. This leads to the Li\'enard--Wiechert potentials, originally due to Li\'enard \cite{Lienard1898} and Wiechert \cite{Wiechert1901}; standard presentations can be found, for example, in Schwinger et al. \cite{Schwinger2018}, Panofsky and Phillips \cite{Panofsky1975}, and Jackson \cite{Jackson1998nia}. A characteristic feature of these potentials is their dependence on the retarded time, which is determined implicitly by both the observation point and the trajectory of the source. An ingenious way to derive the Li\'enard--Wiechert fields is to use a manifestly
Lorentz-covariant derivation of the electromagnetic field produced by an arbitrarily moving charge. The essential idea is simple: at the retarded point on the particle's worldline, one may pass to the instantaneous rest frame of the charge, where the field takes the Coulomb form. The field in the original inertial frame is then obtained by applying the appropriate Lorentz transformation.\footnote{This Lorentz-transformation-based viewpoint goes back to remarks of Minkowski
\cite{Minkowski1909}; see also the discussion in Landau--Lifshitz
\cite{LandauLifshitz1975}, the covariant derivation in \cite[Sec.~23.2.3]{Zangwill2013}, and the detailed note of McDonald
\cite{McDonaldLW2017}. Closely related derivations and discussions may be found
in Fitzpatrick \cite{Fitzpatrick2006}, Strel'tsov
\cite{Streltsov1993}, and Padmanabhan \cite{Padmanabhan:2008gr}. For textbook discussions of retarded potentials, radiation
from moving charges, and the physical content of the Li\'enard--Wiechert fields,
we refer in particular to Griffiths \cite{Griffiths2017}, as well as to
Jackson \cite{Jackson1998nia}, Panofsky--Phillips \cite{Panofsky1975},
Schwinger et al. \cite{Schwinger2018}, Rohrlich \cite{Rohrlich2007}, and the
Feynman lectures \cite{Feynman1964}. Minkowski emphasized the significance of
this example, describing the Li\'enard--Wiechert potentials as
``\textit{perhaps the most striking example}'' of the power of Lorentz
transformations \cite{Minkowski1952}.}



Building on this
AdS perspective, \cite{Duary:2024kxl} further showed that the large-\(L\) limit
of the corresponding AdS field strengths reproduces the familiar antipodal
matching relation near spatial infinity.  Thus the AdS construction provides a
finite-radius geometric origin for the flat-space matching of Coulombic data,
with the usual antipodal relation emerging only after taking the appropriate
large-radius limit.

We revisit this problem from a complementary geometric perspective and to reorganize the derivation in a way that makes the underlying physics more transparent. A basic expectation in the flat-space limit of AdS/CFT is that boundary correlators should reorganize themselves into bulk scattering amplitudes. This idea goes back to the earliest developments of the correspondence~\cite{Polchinski:1999ry,Susskind:1998vk,Giddings:1999qu,Giddings:1999jq}, and has since been developed through a number of related proposals for extracting or characterizing flat-space $\mathcal{S}$-matrix elements from CFT data~\cite{Okuda:2010ym,Penedones:2010ue,Fitzpatrick:2011jn,Maldacena:2015iua,Hijano:2019qmi,Li:2021snj,Duary:2023gqg,Fontanella:2025tbs, Bekaert:2026cib}. More recently, renewed attention has been devoted to the flat-space limit of \emph{gapped} non-gravitational quantum field theories in AdS, where one can study scattering observables in a controlled setting and relate them directly to AdS correlators and bootstrap data~\cite{Dubovsky:2017cnj,Carmi:2018qzm,Komatsu:2020sag,Antunes:2021abs,Gadde:2022ghy,Cordova:2022pbl,Ankur:2023lum, Duary:2022pyv, Banerjee:2022oll, Duary:2022afn, Banerjee:2024yiq, Fernandes:2023xim,Paulos:2016fap}.

The guiding idea is that in AdS the natural analogue of uniform inertial motion in Minkowski space is motion along a timelike geodesic. In flat-space, the field of a uniformly moving charge is related to the static Coulomb field because inertial worldlines are related by Poincar\'e transformations. In AdS, the corresponding statement is that a freely moving charge follows a timelike geodesic \(\Gamma\), and the associated electromagnetic field should therefore be understood as the Coulomb solution viewed in a frame centered on \(\Gamma\). This interpretation is closely related to the general construction of exact Fermi coordinates for timelike geodesic observers in AdS \cite{Klein:2009is}, and is further motivated by the broader usefulness of geodesic-adapted coordinates in isolating physical observables in AdS \cite{Chu:2019ben}.

\begin{figure}[H]
\centering
\begin{tikzpicture}[
    x=1cm,y=1cm,
    >=Latex,
    panel/.style={draw, rounded corners=3pt, very thick, fill=gray!4},
    boundary/.style={thick},
    worldline/.style={very thick},
    nullray/.style={dashed, thin},
    flow/.style={->, very thick},
    lab/.style={font=\small},
    slabel/.style={font=\scriptsize},
    every node/.style={align=center}
]

\begin{scope}[shift={(0,0)}]

  \draw[panel] (-2.7,-2.3) rectangle (2.7,2.4);
  \node[lab] at (0,2.0) {Static problem in global \(\mathrm{AdS}\)};

  \draw[boundary] (0,1.45) ellipse (1.55 and 0.28);
  \draw[boundary] (-1.55,-1.45) -- (-1.55,1.45);
  \draw[boundary] (1.55,-1.45) -- (1.55,1.45);
  \draw[boundary] (0,-1.45) ellipse (1.55 and 0.28);

  \draw[worldline,blue!75!black] (0,-1.15) -- (0,1.15);
  \fill (0,0) circle (1.4pt);
  \node[lab,right=2pt] at (0,0) {$Q$};

  \node[slabel] at (0,-1.98) {$\rho=0$};
  \node[slabel,rotate=90] at (-1.95,0) {$\rho=\frac{\pi}{2}$};
  \node[slabel] at (1.92,1.05) {$\tau$};

\end{scope}

\begin{scope}[shift={(9.0,0)}]

  \draw[panel] (-2.9,-2.3) rectangle (2.9,2.4);
  \node[lab] at (0,2.0) {Geodesic-adapted frame};

  \draw[boundary] (0,1.45) ellipse (1.65 and 0.28);
  \draw[boundary] (-1.65,-1.45) -- (-1.65,1.45);
  \draw[boundary] (1.65,-1.45) -- (1.65,1.45);
  \draw[boundary] (0,-1.45) ellipse (1.65 and 0.28);

  \draw[worldline,red!75!black] (0,-1.15) -- (0,1.15);
  \fill (0,0) circle (1.2pt);

  \draw[nullray] (0,0) -- (-1.55,1.00);
  \draw[nullray] (0,0) -- (1.55,1.00);
  \draw[nullray] (0,0) -- (-1.55,-1.00);
  \draw[nullray] (0,0) -- (1.55,-1.00);

  \fill (-1.52,1.00) circle (1pt);
  \fill (1.52,-1.00) circle (1pt);

  \node[slabel,left=2pt] at (-1.55,1.00) {$\tau=+\frac{\pi}{2}$};
  \node[slabel,right=2pt] at (1.55,-1.00) {$\tau=-\frac{\pi}{2}$};

\end{scope}

\draw[flow] (3.1,0.45) -- (5.9,0.45);
\node[lab] at (4.5,1.0) {adapt to\\ timelike geodesic \(\Gamma\)};
\node[slabel] at (4.5,-0.2) {``uniform motion'' \\ \(=\) geodesic motion};

\draw[flow] (0,-2.8) -- (0,-3.8);
\draw[flow] (9.0,-2.8) -- (9.0,-3.8);

\node[lab] at (0,-4.4) {solve the static\\ Coulomb problem once};
\node[lab] at (9.0,-4.4) {rewrite in arbitrary global coordinates\\ using embedding invariants, then take \(L\to\infty\)};

\node[slabel] at (9.0,-1.85) {null rays select the boundary regions \(\mathscr{I}^\pm\)};

\end{tikzpicture}
\caption{Geometric strategy of the derivation. The moving-charge problem in \(\mathrm{AdS}_4\) is reduced to the static Coulomb problem by introducing coordinates adapted to the source geodesic \(\Gamma\). In the adapted frame, the field is again central; the nontrivial step is to rewrite it in arbitrary global coordinates using embedding-space invariants. The null rays from the geodesic pick out the distinguished boundary regions near \(\tau=\pm \pi/2\), whose large-\(L\) limit yields future and past null infinity and leads to antipodal matching.}
\label{fig:strategy_improved}
\end{figure}
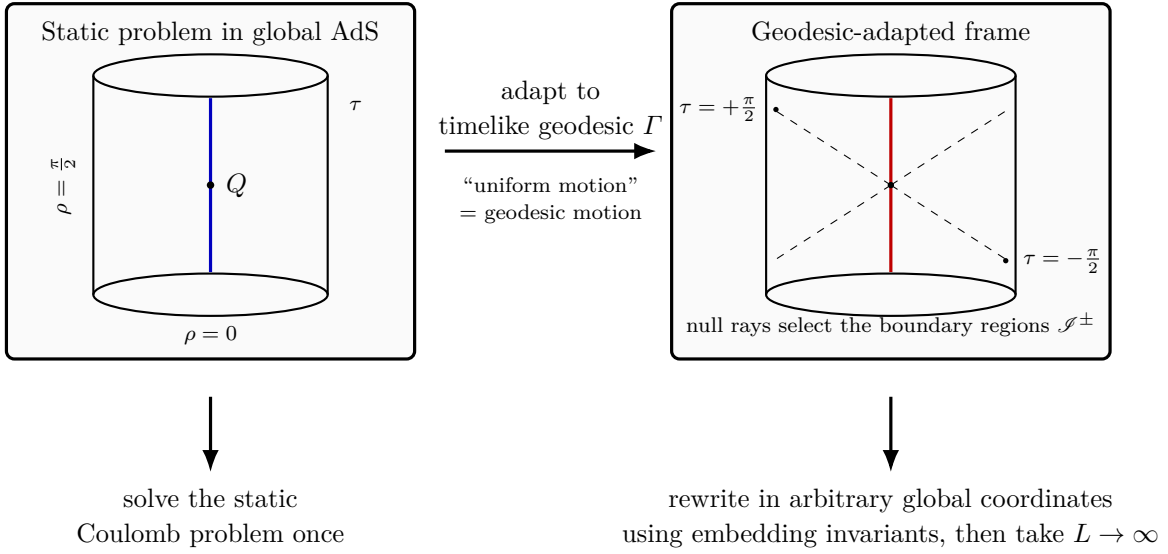

Our strategy is therefore not to generate the moving solution by explicitly boosting a known answer, but rather to reformulate the problem geometrically. We first solve the static Maxwell problem for a charge placed at the center of global \(\mathrm{AdS}_4\). We then introduce coordinates adapted to an arbitrary timelike geodesic \(\Gamma\), so that in the adapted frame the source once again sits at the center. In this frame the field is therefore the same static Coulomb field. The nontrivial step is then to rewrite this central solution in arbitrary global coordinates. Because AdS is a maximally symmetric spacetime, this rewriting is naturally expressed in terms of embedding-space invariants, in the spirit of the invariant geometric constructions developed in \cite{Allen:1985wd}. The logic of the method is summarized in Figure \ref{fig:strategy_improved}. The left panel represents the static problem at the center of global AdS, where the source sits at \(\rho=0\) and the field is the ordinary Coulomb field adapted to AdS geometry. The right panel represents the same physics after recentering the geometry on an arbitrary timelike geodesic \(\Gamma\). In this geodesic-adapted frame, the source is again at the center, so the field remains the static Coulomb solution; what changes is only the geometric identification of the observer and the boundary.

This perspective makes two important facts especially transparent. First, the AdS analogue of the Li\'enard--Wiechert field is fundamentally a geodesic observable: the field of a charge in ``uniform motion'' is simply the static Coulomb field expressed in coordinates centered on the appropriate timelike geodesic. Second, the origin of antipodal matching becomes geometric. In the embedding-space description of AdS, boundary points are represented by projective null rays, and opposite null directions are naturally identified by the antipodal map. In the large-\(L\) limit, the null rays emitted from the source geodesic reach the boundary near the distinguished regions \(\tau=\pm \pi/2\), which become the flat-space null infinities \(\mathscr{I}^\pm\). The antipodal matching relation then emerges as the flat-space remnant of this projective null structure, in agreement with the flat-space analysis of \cite{Strominger:2017zoo}.

With this interpretation in mind, our derivation proceeds in the following steps
\begin{enumerate}[label=(\arabic*)]
    \item We solve Maxwell's equations for a static point charge at the center of global \(\mathrm{AdS}_4\).
    \item Given an arbitrary timelike geodesic \(\Gamma\), we introduce coordinates adapted to \(\Gamma\), in which the worldline again sits at the center.
    \item In this adapted frame the field is exactly the static Coulomb solution.
    \item We then rewrite the resulting field in arbitrary global coordinates using AdS embedding-space invariants.
    \item Finally, we take the AdS boundary limit and then the large-\(L\) flat-space limit, recovering the standard antipodal matching relation.
\end{enumerate}

\paragraph{Antipodal matching and the infrared structure of scattering.}
A central lesson from the modern understanding of infrared physics is that the asymptotic data at past and future null infinity should not be glued at spatial infinity by matching fields at the same angular point. Already in classical electrodynamics, the Li\'enard--Wiechert field of a uniformly moving charge has different limiting values at \(\mathscr I^+_-\) and \(\mathscr I^-_+\) at fixed angle. The Lorentz-invariant gluing condition instead identifies the leading radial electric field at antipodal points,
\begin{equation}
\label{eq:flat_antipodal_matching_intro}
\left. \lim_{r\to\infty} r^2 F_{ru}(u,\hat x) \right|_{\mathscr I^+_-}
=
\left. \lim_{r\to\infty} r^2 F_{rv}(v,-\hat x) \right|_{\mathscr I^-_+}. \nonumber
\end{equation}
This antipodal matching condition is the basic boundary condition underlying the infinite set of conservation laws generated by large gauge transformations at null infinity, and it plays a central role in the infrared triangle relating soft theorems, asymptotic symmetries, and memory effects \cite{Strominger:2017zoo}. In QED, the same large-gauge charges imply selection rules for the \(S\)-matrix: generic scattering processes do not preserve a single Fock vacuum, but rather induce transitions among the infinitely many degenerate vacua labelled by soft photon zero modes. When these vacuum transitions are properly included, the resulting amplitudes are infrared finite and are equivalent to the Faddeev--Kulish construction \cite{Kapec:2017tkm} (see also \cite{Duary:2022afn} for AdS correction to the Faddeev--Kulish state).\footnote{See also \cite{Hannesdottir:2019umk,Hannesdottir:2019opa} for a complementary formulation of infrared-finite scattering in theories with massless particles, in which the universal asymptotic dynamics are separated from the hard process and incorporated into a modified, or hard, $\mathcal{S}$-matrix.}
The gravitational analogue is provided by BMS supertranslation symmetry, where amplitudes conserving the supertranslation charge are precisely the corresponding Faddeev--Kulish amplitudes and are therefore infrared finite \cite{Choi:2017ylo}. Thus antipodal matching is not merely a convenient asymptotic prescription; it is the classical gluing condition behind asymptotic charge conservation, soft vacuum transitions, and infrared-finite scattering. In the present work, we reinterpret this structure from the viewpoint of global \(\mathrm{AdS}_4\), showing that the flat-space antipodal matching condition arises as the null-fringe limit of an exact finite-radius antipodal covariance of the AdS Li\'enard--Wiechert field.


\subsection{Summary of main results and conceptual strategy}
\label{subsec:intro-main-results}
The purpose of this paper is to give a geometric reformulation of Li\'enard--Wiechert fields in flat-space and in $\mathrm{AdS}_4$, and to use this reformulation to clarify the origin of antipodal matching. The central idea is simple but powerful: the field of a uniformly moving charge should not be viewed as a fundamentally new solution, but rather as the ordinary Coulomb field expressed in coordinates adapted to the source worldline. In flat-space, the relevant worldline is a timelike geodesic of Minkowski space; in AdS, the natural analogue of uniform inertial motion is motion along a timelike geodesic of the AdS hyperboloid. This point of view allows the entire problem to be reorganized around the geometry of geodesics and the coordinate systems naturally centered on them, rather than around explicit boosts or isometry transformations of already-known answers. The result is a derivation in which both the Li\'enard--Wiechert field and the associated antipodal matching relation emerge directly from a rest-frame Coulomb seed together with the geometry of the underlying spacetime.

Our analysis proceeds in two stages. We first revisit the flat-space problem from a deliberately geometric perspective. Instead of starting from the standard Li\'enard--Wiechert potential in inertial coordinates, we introduce coordinates $(T,R,\Omega)$ adapted to the timelike geodesic followed by a uniformly moving charge. In these coordinates the charge is static at $R=0$, and the gauge field is simply the ordinary Coulomb solution
$$
A=-\frac{q}{4\pi R}\,dT.
$$
The nontrivial step is then to rewrite this solution back in arbitrary inertial coordinates. Carrying this out yields the standard field of a uniformly moving charge, but now interpreted as the Coulomb field of the source seen in a frame not centered on its worldline. This perspective makes the physics particularly transparent: the anisotropy of the field at null infinity is not a radiative effect, but the asymptotic imprint of a fundamentally Coulombic field after it is embedded into an arbitrary inertial slicing. From this form of the field we recover the leading Coulombic data at $\mathscr I^+_-$ and $\mathscr I^-_+$ and show directly that their equality requires the antipodal map on the celestial sphere. In this way, the flat-space matching law is derived not as an additional boundary prescription, but as a geometric continuity statement along the null generators of conformal infinity.

We then generalize the same construction to global $\mathrm{AdS}_4$. We begin by solving the static Maxwell problem for a point charge placed at the center of AdS, obtaining the Coulomb seed
$$
A=-\frac{q}{4\pi}\cot\rho\,d\tau,
\qquad
F=\frac{q}{4\pi}\csc^2\rho\,d\rho\wedge d\tau.
$$
This is the unique static, spherically symmetric solution compatible with the local flat-space normalization near the AdS center. The key observation is that once this central solution is known, the moving-charge problem does not require solving Maxwell's equations again from scratch. Instead, one introduces coordinates $(T,R,\Theta,\Phi)$ adapted to an arbitrary timelike AdS geodesic $\Gamma$, so that in the adapted frame the charge is again at rest at $R=0$. In this frame the field is therefore once more the static Coulomb solution,
$$
A=-\frac{q}{4\pi}\cot R\,dT,
\qquad
F=\frac{q}{4\pi}\csc^2R\,dR\wedge dT.
$$
The dynamical content of the moving solution is thus entirely encoded in the reconstruction of the adapted coordinates $(T,R)$ in terms of the original global coordinates $(\tau,\rho,\Omega)$. Because AdS is maximally symmetric, this reconstruction can be carried out invariantly in embedding space, where AdS is realized as the hyperboloid $X\cdot X=-L^2$ in $\mathbb R^{3,2}$. Using the embedding-space data $(P,U)$ that characterize the source geodesic, we derive explicit invariant formulas for the adapted time and radius in terms of the inner products $P\cdot X$ and $U\cdot X$. This gives a closed-form geometric expression for the AdS Li\'enard--Wiechert field in arbitrary global coordinates.

One of the main technical results of the paper is the explicit global-coordinate expression for the gauge-invariant field-strength component $F_{\rho\tau}(\tau,\rho,\hat x)$ associated with a charge moving on a timelike geodesic through the AdS origin. Writing the source velocity as $\vec\beta=\beta \hat p$ and defining $\mu=\hat p\cdot \hat x$, we find
$$
F_{\rho\tau}(\tau,\rho,\hat x)
=
\frac{q\gamma\bigl(\sin\rho-\beta\mu\sin\tau\bigr)}
{4\pi\Bigl(\cos^2\tau-\cos^2\rho+\gamma^2\bigl(\sin\tau-\beta\sin\rho\,\mu\bigr)^2\Bigr)^{3/2}},
$$
or equivalently
$$
F_{\rho\tau}(\tau,\rho,\hat x)
=
\frac{q\gamma\bigl(\sin\rho-\sin\tau\,\vec\beta\cdot\hat x\bigr)}
{4\pi\Bigl(\cos^2\tau-\cos^2\rho+\gamma^2\bigl(\sin\tau-\sin\rho\,\vec\beta\cdot\hat x\bigr)^2\Bigr)^{3/2}}.
$$
This is the exact AdS analogue of the Coulombic part of the flat-space Li\'enard--Wiechert field. It is a genuine bulk quantity on the global AdS cylinder and reduces correctly both to the static Coulomb seed when $\beta\to 0$ and to the expected flat-space behavior near the AdS center in the large-$L$ limit.

A second main result is that the bulk AdS solution already satisfies an exact antipodal covariance before any boundary or flat-space limit is taken. In global coordinates the AdS antipodal map acts as
$$
\mathscr A:\qquad
\tau\to \tau\pm\pi,\qquad
\rho\to\rho,\qquad
\Omega\to\Omega_A,
$$
where $\Omega_A$ is the antipodal point on $S^2$. We show directly from the explicit bulk expression that
$$
F_{\rho\tau}(\tau\pm\pi,\rho,\Omega_A)=F_{\rho\tau}(\tau,\rho,\Omega)
$$
for arbitrary bulk radius $\rho$. Thus the bulk field obeys a stronger statement than the usual flat-space matching law: it is exactly covariant under the finite-radius AdS antipodal isometry. This is a purely geometric consequence of the embedding-space inversion $X^A\to -X^A$, which preserves the AdS hyperboloid and induces the above transformation in global coordinates. In this sense, antipodal matching is already present in the AdS bulk, and the flat-space relation should be viewed as a descendant of this stronger finite-radius statement.

To connect with flat-space infrared physics, however, one must isolate the asymptotic data that survive the large-$L$ limit. Global AdS makes this structure especially transparent. Radial null rays emitted from the AdS center at $\tau=0$ reach the conformal boundary at the distinguished times $\tau=\pm \pi/2$. The corresponding thin boundary neighborhoods are the future and past null fringes of global AdS, and in the large-$L$ limit they become the avatars of $\mathscr I^\pm$. Evaluating the bulk field in these limits, we obtain
$$
\lim_{\substack{\rho\to\pi/2\\ \tau\to+\pi/2}}
F_{\rho\tau}(\tau,\rho,\hat x)
=
\frac{q}{4\pi\gamma^2\bigl(1-\vec\beta\cdot\hat x\bigr)^2},
$$
and
$$
\lim_{\substack{\rho\to\pi/2\\ \tau\to-\pi/2}}
F_{\rho\tau}(\tau,\rho,\hat x)
=
\frac{q}{4\pi\gamma^2\bigl(1+\vec\beta\cdot\hat x\bigr)^2}.
$$
These are the precise AdS null-fringe analogues of the leading Coulombic data at $\mathscr I^+_-$ and $\mathscr I^-_+$. They are not equal at the same angle, but they are related by the antipodal map,
$$
\lim_{\substack{\rho\to\pi/2\\ \tau\to+\pi/2}}
F_{\rho\tau}(\tau,\rho,\hat x)
=
\lim_{\substack{\rho\to\pi/2\\ \tau\to-\pi/2}}
F_{\rho\tau}(\tau,\rho,-\hat x).
$$
Thus the usual flat-space matching law is recovered as the restriction of the exact AdS antipodal covariance to the boundary null fringes, followed by the large-$L$ limit. This shows that the familiar matching condition at $\mathscr I^+_-$ and $\mathscr I^-_+$ is the large-radius remnant of the projective null structure of the AdS boundary.

The method developed in this paper is important for several related reasons. First, it replaces an isometry-based or boost-based derivation by a viewpoint that is manifestly geometric and conceptually uniform in both flat-space and AdS. In this formulation, the field of a moving charge is never treated as a fundamentally distinct solution. Rather, it is always the same Coulomb seed, expressed in coordinates centered on the appropriate timelike geodesic and then rewritten in a more general frame. This shift of viewpoint is useful because it disentangles what is intrinsic from what is purely kinematical. The intrinsic content is the local Coulombic field seen in the rest frame of the source, while the nontrivial angular dependence that appears at null infinity is entirely a consequence of how that rest-frame solution is embedded into a global spacetime slicing. In this sense, the anisotropy of the asymptotic field is not itself a new dynamical effect, but the geometric image of a simple radial field under a change of frame adapted to the source trajectory.

Second, the construction separates the problem into two logically distinct layers. One is the local rest-frame problem, which is simple, universal, and largely independent of the global details of the motion. The other is the reconstruction problem, namely how the adapted coordinates associated with the source worldline are expressed in a chosen global coordinate system. All of the nontrivial kinematics of the moving solution are concentrated in this second step. 

Third, the geodesic-centered viewpoint makes the origin of antipodal matching especially transparent. In flat-space, the matching law is often stated as an asymptotic condition imposed on fields near spatial infinity. In the present formulation, however, it appears instead as a direct consequence of the fact that the leading Coulombic field is continuous along the null generators of conformal infinity. The necessity of the antipodal map is then not mysterious: the relevant generator enters from $\mathscr I^-_+$ and exits at the antipodal point of $\mathscr I^+_-$. In AdS, the same logic survives in a sharper form. The bulk field already satisfies an exact antipodal covariance at finite radius, and the familiar flat-space matching law emerges only after restricting this stronger statement to the distinguished null-fringe regions near $\tau=\pm \pi/2$ and then taking the large-$L$ limit. 

A further virtue of the present formulation is that it interfaces naturally with the flat-space limit of AdS/CFT. The global AdS cylinder singles out the two null-fringe regions near $\tau=\pm \pi/2$ as the natural carriers of asymptotic in- and out-data, and the explicit Coulombic profiles obtained here indicate how the infrared information of Minkowski scattering should be encoded in boundary observables supported on those strips. This point of view also fits naturally with recent celestial and Carrollian approaches to the flat-space limit \cite{deGioia:2024yne,deGioia:2022fcn,Bagchi:2023fbj,Alday:2024yyj,Navarro:2025xln, Bagchi:2026emg}. In those frameworks, the boundary regions near $\tau=\pm \pi/2$ play a distinguished role because they are precisely the regions from which flat-space in/out observables emerge after the large-radius limit. The present construction provides a concrete classical example of that general picture. The future and past null fringes are not auxiliary regions introduced for convenience; they are selected geometrically by radial null propagation from the bulk and therefore carry the correct asymptotic Coulombic data. The recovery of antipodal matching from these regions strongly suggests that the infrared structure of flat-space scattering is already encoded in the geometry of the AdS boundary at finite radius, before one takes the flat-space limit. In that sense, the Li\'enard--Wiechert field provides a particularly clean probe of how AdS data reorganize into celestial or Carrollian observables \cite{Duary:2022onm, Duary:2024fii, Duary:2024cqb}.
In summary, the main results of this paper may be viewed as establishing a unified geometric picture of Li\'enard--Wiechert fields in flat-space and in $\mathrm{AdS}_4$. We first gave a purely geometric derivation of the flat-space Li\'enard--Wiechert field and the corresponding antipodal matching law by introducing coordinates adapted to the source timelike geodesic. We then solved the static Coulomb problem in global $\mathrm{AdS}_4$, introduced geodesic-adapted coordinates for an arbitrary timelike AdS geodesic, and reconstructed the moving field in arbitrary global coordinates using embedding-space invariants. This led to an explicit closed-form expression for $F_{\rho\tau}(\tau,\rho,\hat x)$, an exact bulk antipodal covariance under the AdS antipodal map, and a derivation of the standard flat-space antipodal matching relation from the future and past null-fringe limits in the large-$L$ regime. The broader significance of the construction is that it identifies the AdS Li\'enard--Wiechert field as a genuinely geodesic observable and shows that the infrared matching law of flat-space electrodynamics is best understood as the asymptotic shadow of a stronger geometric structure already present at finite AdS radius.

There is also a complementary image-charge viewpoint on this construction. In flat-space, conformal compactification turns a spatial slice into a compact $S^3$, and Gauss's law then represents the Coulomb field of an isolated charge together with a compensating image singularity at spatial infinity $i^0$. This image charge is not an additional physical source, but the compactified representation of the Coulomb flux through infinity. It explains why smoothness at $i^0$ is too strong a requirement, while the meaningful data are the leading Coulombic fields transported along null generators, giving the antipodal matching condition. In AdS, the analogous statement is boundary-condition dependent: after conformal rescaling, a constant global-time slice is a hemisphere of $S^3$, and the static potential $\Phi=(q/4\pi)\cot\rho$ is the Dirichlet Green function on this hemisphere. Doubling the hemisphere across the AdS boundary represents this field as that of a charge at the AdS center together with an opposite image charge in the reflected copy. Thus the image-charge picture explains how the Coulomb seed realizes the AdS boundary condition, while the antipodal covariance explains how the corresponding future and past null-fringe data are related. In geodesic-adapted coordinates the same interpretation applies to the moving solution: the AdS Li\'enard--Wiechert field is the Coulomb field of the charge, together with its boundary image, rewritten in a frame centered on the source geodesic.

The paper is organized as follows.  The present summary is
Subsection~\ref{subsec:intro-main-results}.  Section~\ref{sec:global-ads-flat-limit}
reviews global AdS and its flat-space limit, while
Subsection~\ref{subsec:witten-diagrams-bulk-point} relates the distinguished
boundary times to the bulk-point limit.  Section~\ref{sec:flat-geodesic-antipodal}
develops the flat-space derivation: Subsections~\ref{subsec:flat-geodesic-coordinates}--\ref{subsec:flat-field-inertial}
build and rewrite the geodesic-centered Coulomb field,
Subsections~\ref{subsec:flat-future-null-infinity} and
\ref{subsec:flat-past-null-infinity} extract the two null-infinity limits, and
Subsections~\ref{subsec:flat-antipodal-matching} and
\ref{subsec:flat-conformal-image} derive the matching law and the compactified
image-charge interpretation.  Section~\ref{sec:ads-null-fringe-times}
identifies the AdS null fringes, and Section~\ref{sec:maxwell-global-ads},
with Subsection~\ref{subsec:ads-static-coulomb-center}, solves the static
Maxwell problem.  Section~\ref{sec:ads-geodesic-coordinates} develops the AdS
geodesic construction through Subsections~\ref{subsec:ads-geodesics-embedding},
\ref{subsec:ads-geodesic-uniform-motion}, \ref{subsec:ads-geodesic-adapted-coordinates},
\ref{subsec:ads-invariant-reconstruction}, and
\ref{subsec:ads-explicit-origin-geodesic}.  Section~\ref{sec:ads-field-strength-antipodal},
together with Subsections~\ref{subsec:ads-global-frhotau} and
\ref{subsec:ads-antipodal-matching}, derives the global field strength and its
AdS antipodal matching.  Section~\ref{sec:ads-antipodal-map} explains the
embedding-space origin of the antipodal map, Section~\ref{sec:ads_image_charge}
develops the AdS image-charge picture, and Section~\ref{sec:conclusions}
concludes.  Appendix~\ref{app:image-charge-spatial-infinity} gives the
flat-space image-charge analysis.

\section{Global AdS and the flat-space limit}
\label{sec:global-ads-flat-limit}

We begin with the standard realization of Lorentzian $\mathrm{AdS}_4$ as the quadric
\begin{equation}
-(X^1)^2-(X^2)^2+(X^3)^2+(X^4)^2+(X^5)^2=-L^2
\end{equation}
embedded in the ambient space $\mathbb{R}^{3,2}$ with metric
\begin{equation}
\eta_{AB}=\mathrm{diag}(-1,-1,+1,+1,+1).
\end{equation}

A convenient global parametrization of the hyperboloid is
\begin{align}
X^1 &= \frac{L\cos\tau}{\cos\rho}, &
X^2 &= \frac{L\sin\tau}{\cos\rho}, \\
X^3 &= L\tan\rho\,\sin\theta\cos\phi, &
X^4 &= L\tan\rho\,\sin\theta\sin\phi, &
X^5 &= L\tan\rho\,\cos\theta,
\end{align}
for which the induced metric takes the familiar form
\begin{equation}
\label{globalmetric}
ds^2=\frac{L^2}{\cos^2\rho}\Bigl(-d\tau^2+d\rho^2+\sin^2\rho\,d\Omega_2^2\Bigr),
\qquad 0\le \rho<\frac{\pi}{2}.
\end{equation}
The conformal boundary lies at $\rho=\pi/2$. After stripping off the divergent Weyl factor $L^2/\cos^2\rho$, one obtains the boundary cylinder
\begin{equation}
ds^2_{\partial \mathrm{AdS}}=-d\tau^2+d\Omega_2^2,
\end{equation}
namely $\mathbb{R}_\tau\times S^2$.

\begin{figure}[H]
\centering
\begin{tikzpicture}[scale=1.15,>=Latex]

  \draw[thick] (-2.2,-3) -- (-2.2,3);
  \draw[thick] (2.2,-3) -- (2.2,3);

  \draw[thick] (0,3) ellipse (2.2 and 0.45);
  \draw[thick] (0,-3) ellipse (2.2 and 0.45);

  \draw[dashed] (0,-3) -- (0,3);

  \fill (0,0) circle (1.5pt);
  \node[right] at (0.05,0.15) {$\small \rho=0,\ \tau=0$};

  \draw[very thick,blue!70!black] (0,0) -- (-2.2,2.2);
  \draw[very thick,blue!70!black] (0,0) -- (2.2,2.2);
  \draw[very thick,blue!70!black] (0,0) -- (-2.2,-2.2);
  \draw[very thick,blue!70!black] (0,0) -- (2.2,-2.2);

  \fill[red!15] (-2.2,2.05) rectangle (2.2,2.35);
  \fill[blue!15] (-2.2,-2.35) rectangle (2.2,-2.05);

  \node[left] at (-2.35,2.2) {$\tau=+\frac{\pi}{2}$};
  \node[left] at (-2.35,-2.2) {$\tau=-\frac{\pi}{2}$};

  \node at (0,3.10) {$\text{future null fringe}$};
  \node at (0,-3.20) {$\text{past null fringe}$};

  \node[rotate=90] at (-2.55,0) {$\rho=\frac{\pi}{2}$};
  \node[rotate=90] at (2.55,0) {$\rho=\frac{\pi}{2}$};

  \node at (0,3.70) {$\partial \mathrm{AdS}_4 \cong \mathbb{R}_\tau \times S^2$};

  \draw[->] (2.8,-2.7) -- (2.8,2.7);
  \node[right] at (2.8,2.7) {$\tau$};

  \draw[->] (0,0) -- (1.4,0);
  \node[below] at (0.9,0) {$\rho$};

\end{tikzpicture}
\caption{Global AdS as a cylinder. The conformal boundary sits at $\rho=\pi/2$. Radial null geodesics emitted from the center at $\tau=0$ reach the boundary at $\tau=\pm \pi/2$. In the large-$L$ limit, thin strips around these two times become the avatars of future and past null infinity, $\mathscr{I}^\pm$.}
\label{fig:globalAdsCylinder}
\end{figure}
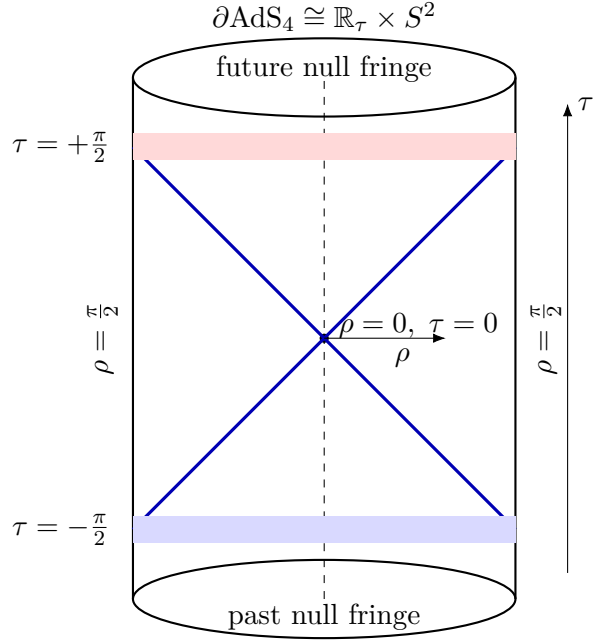
Global coordinates are particularly well suited to the present problem. First, they make the causal structure of AdS manifest and therefore provide a natural arena for discussing null and timelike geodesics. Second, they exhibit the flat-space limit in a geometrically transparent way. To zoom into the central scattering region, one introduces the rescaled coordinates
\begin{equation}
\tau=\frac{t}{L},
\qquad
\tan\rho=\frac{r}{L}.
\end{equation}
Substituting these relations into \eqref{globalmetric}, one finds
\begin{equation}
ds^2
=
-\Bigl(1+\frac{r^2}{L^2}\Bigr)dt^2
+
\frac{dr^2}{1+r^2/L^2}
+
r^2 d\Omega_2^2.
\end{equation}
Hence, for fixed $(t,r,\theta,\phi)$ and $L\to\infty$,
\begin{equation}
ds^2 \xrightarrow[L\to\infty]{} -dt^2+dr^2+r^2 d\Omega_2^2,
\end{equation}
so the geometry reduces locally to Minkowski space. In other words, the flat-space limit probes a neighborhood of the AdS center whose size is small compared with the curvature scale $L$.

A further virtue of global coordinates is that they make the distinguished role of the times $\tau=\pm \pi/2$ immediate. Indeed, for radial null motion one has $d\Omega_2=0$ and $ds^2=0$, so \eqref{globalmetric} implies
\begin{equation}
d\tau=\pm d\rho.
\end{equation}
A null geodesic emitted from the center $\rho=0$ at $\tau=0$ therefore reaches the boundary $\rho=\pi/2$ at
\begin{equation}
\tau=\pm \frac{\pi}{2}.
\end{equation}
Thus the thin boundary strips around $\tau=+\pi/2$ and $\tau=-\pi/2$ are the global AdS precursors of future and past null infinity in the large-$L$ limit. This observation is central for the flat-space scattering interpretation: boundary data localized near these two null fringes are precisely the data that survive as asymptotic in- and out-data in the Minkowski limit. It is in this sense that the regions near $\tau=\pm\pi/2$ play the role of $\mathscr{I}^\pm$, and it is this geometric structure that underlies the emergence of antipodal matching in the flat-space limit. See Figure \ref{fig:globalAdsCylinder}.

\subsection{Witten diagrams at the bulk point}
\label{subsec:witten-diagrams-bulk-point}

The distinguished boundary times \(\tau=\pm \pi/2\) admit an important second interpretation. Besides being the arrival times of radial null geodesics emitted from the AdS center, they are also precisely the boundary slices naturally associated with the Lorentzian bulk-point limit of AdS correlators \cite{Maldacena:2015iua,Lam:2017ofc}. In a local bulk description, a Witten diagram develops a bulk-point singularity when the boundary insertions can be connected to a single bulk point \(Y_0\) by null geodesics, and when positive frequencies can be assigned to those null rays so that momentum is conserved at \(Y_0\). In embedding-space language, if the boundary insertions are represented by null rays \(P_a\), the bulk-point conditions are
\begin{equation}
Y_0\cdot P_a = 0,
\qquad
\sum_a \omega_a P_a = 0,
\qquad
\omega_a>0.
\end{equation}
These are the Lorentzian Landau conditions for a pointlike bulk scattering process \cite{Maldacena:2015iua}.

Now choose the reference bulk point to be the AdS center. In the embedding description this may be taken to be
\begin{equation}
Y_0=(L,0,0,0,0).
\end{equation}
The null-separation condition \(Y_0\cdot P_a=0\) then forces the relevant boundary insertions to lie on the two constant-time slices
\begin{equation}
\tau=\pm \frac{\pi}{2}.
\end{equation}
Thus the same null-fringe regions that emerged above from the causal structure of global AdS are also the regions singled out by the bulk-point limit. In this sense, the future and past null fringes are not auxiliary constructs: they are exactly the boundary loci on which the in- and out-data of the local bulk scattering process are supported \cite{Maldacena:2015iua}.

Near the bulk point, the dominant contribution to the Witten diagram comes from an arbitrarily small neighborhood of \(Y_0\), where the AdS geometry becomes approximately flat. In this regime the boundary null rays take the form
\begin{equation}
P_a=(0,\pm q_a^\mu),
\qquad
q_a q_{a}=0,
\end{equation}
so that the conservation law reduces to the ordinary flat-space null-momentum conservation condition
\begin{equation}
\sum_{\rm out}\omega_a q_a^\mu-\sum_{\rm in}\omega_a q_a^\mu=0.
\end{equation}
Correspondingly, the AdS bulk-to-boundary propagators reduce to flat-space conformal primary wavefunctions, and the leading bulk-point contribution to the Witten diagram is governed by the same flat-space amplitude written in the conformal basis \cite{Lam:2017ofc}. In this way, the bulk-point limit provides a concrete mechanism by which Lorentzian AdS correlators reorganize into flat-space scattering data.

This picture is fully consistent with the geometric derivation of the AdS Li\'enard--Wiechert field developed in the present work. Here too, the large-\(L\) limit isolates a small neighborhood of the AdS center together with boundary data localized near \(\tau=\pm \pi/2\). From the bulk-point point of view, these strips carry the in- and out-data of a local bulk scattering process. From the geodesic-centered Coulomb point of view, they carry the leading Coulombic data of the moving charge. The fact that both descriptions single out the same null fringes strongly suggests that the antipodal matching relation obtained below is the classical gauge-field counterpart of the same AdS-to-flat reorganization that underlies the bulk-point limit.

As we will see, the Li\'enard--Wiechert fields in AdS and their antipodal matching fit naturally into this picture. The bulk solution already exhibits an exact antipodal covariance at finite AdS radius, while its restriction to the conformal boundary near \(\tau=\pm \pi/2\) reproduces the standard flat-space antipodal matching law in the large-\(L\) limit. It is therefore natural to regard the AdS Li\'enard--Wiechert field as a particularly simple classical probe of the same null-fringe kinematics that controls the bulk-point regime. The regions selected by bulk-point scattering are precisely those on which the Coulombic infrared data live.

Finally, one should keep in mind the usual caveat: the strict bulk-point singularity is a property of local bulk perturbation theory. Finite \(\alpha'\) effects smooth the singularity, and at finite \(G_N\) one does not expect an exact nonperturbative singular behavior \cite{Maldacena:2015iua}. For the purposes of the present discussion, however, the essential point is kinematical: the bulk-point limit explains why the boundary strips near \(\tau=\pm \pi/2\) are the natural carriers of flat-space scattering data, and our Li\'enard--Wiechert construction shows that the same strips also carry the infrared Coulombic data obeying antipodal matching.

\section{Flat-space geodesic-centered derivation of antipodal matching}
\label{sec:flat-geodesic-antipodal}

In this section we give a purely Minkowskian derivation of the antipodal matching condition for the leading Coulombic component of the electromagnetic field strength. The point of view is deliberately geometric. Rather than beginning with the standard Liénard--Wiechert potential in an inertial frame and then taking its asymptotic limits, we first recast the problem in coordinates adapted to the timelike geodesic followed by a uniformly moving charge. In this adapted frame the source is static and the field is simply the ordinary Coulomb field. The nontrivial step is then to rewrite this field in arbitrary inertial coordinates and study its behavior near null infinity. In this way the familiar angular dependence of the asymptotic electric field emerges directly from the geometry of the source worldline, and the antipodal matching condition appears as the natural continuity statement along the null generators of conformal infinity.

Throughout this section we work in natural units and use the mostly-plus convention
\begin{equation}
\begin{split}
c=\hbar=1,
\qquad
\eta_{\mu\nu}=\mathrm{diag}(-,+,+,+).
\end{split}
\end{equation}

\subsection{Timelike geodesic and geodesic-centered coordinates}
\label{subsec:flat-geodesic-coordinates}

Consider a point charge moving with constant three-velocity $\vec\beta$. Its four-velocity is
\begin{equation}
\begin{split}
u^\mu=\gamma(1,\vec\beta),
\qquad
\gamma=\frac{1}{\sqrt{1-\beta^2}},
\qquad
u^\mu u_\mu=-1,
\end{split}
\end{equation}
and its worldline is
\begin{equation}
\begin{split}
X^\mu(s)=u^\mu s,
\end{split}
\end{equation}
where $s$ is proper time. Since $u^\mu$ is constant, the source follows a timelike geodesic of Minkowski space. The essential observation is that a uniformly moving charge is not intrinsically different from a static one: the distinction is purely frame-dependent. What changes from one inertial observer to another is the slicing of spacetime, not the underlying Coulombic character of the field.

To make this statement explicit, let $x^\mu$ be an arbitrary bulk point and decompose it into components parallel and orthogonal to $u^\mu$. We define the geodesic-centered time coordinate by
\begin{equation}
\begin{split}
T:=-u\cdot x,
\end{split}
\end{equation}
and the orthogonal displacement by
\begin{equation}
\begin{split}
Y^\mu:=x^\mu+(u\cdot x)u^\mu.
\end{split}
\end{equation}
By construction,
\begin{equation}
\begin{split}
u\cdot Y
&=
u\cdot x+(u\cdot x)u^2 \\
&=
u\cdot x-(u\cdot x)=0,
\end{split}
\end{equation}
so $Y^\mu$ lies in the Euclidean hyperplane orthogonal to the source four-velocity. The natural spatial radius from the worldline is therefore
\begin{equation}
\begin{split}
R:=\sqrt{Y^\mu Y_\mu}.
\end{split}
\end{equation}
Using the definition of $Y^\mu$, one immediately finds
\begin{equation}
\begin{split}
R^2
&=
\bigl(x^\mu+(u\cdot x)u^\mu\bigr)\bigl(x_\mu+(u\cdot x)u_\mu\bigr) \\
&=
x^2+(u\cdot x)^2.
\end{split}
\end{equation}
Hence the invariant form of the adapted coordinates is
\begin{equation}
\begin{split}
T=-u\cdot x,
\qquad
R=\sqrt{x^2+(u\cdot x)^2}.
\end{split}
\end{equation}

These coordinates are precisely centered on the source geodesic. Indeed, setting $x^\mu=X^\mu(s)=u^\mu s$ gives
\begin{equation}
\begin{split}
T=-u\cdot(us)=-u^2s=s,
\end{split}
\end{equation}
while
\begin{equation}
\begin{split}
R^2
&=
u^2s^2+(u^2s)^2 \\
&=
(-1)s^2+(-s)^2=0.
\end{split}
\end{equation}
Thus the charge sits at
\begin{equation}
\begin{split}
R=0
\end{split}
\end{equation}
for all $T$. In this sense $(T,R,\Omega)$ are the natural rest-frame coordinates of the moving charge.

The metric also takes a simple form in these variables. Writing
\begin{equation}
\begin{split}
x^\mu=Tu^\mu+Y^\mu,
\qquad
u\cdot Y=0,
\end{split}
\end{equation}
we have
\begin{equation}
\begin{split}
dx^\mu=u^\mu dT+dY^\mu.
\end{split}
\end{equation}
Therefore
\begin{equation}
\begin{split}
ds^2
&=
\eta_{\mu\nu}dx^\mu dx^\nu \\
&=
u^2 dT^2+2(u\cdot dY)dT+dY^2.
\end{split}
\end{equation}
Since $u^2=-1$ and $u\cdot dY=0$, this reduces to
\begin{equation}
\begin{split}
ds^2=-dT^2+dY^2.
\end{split}
\end{equation}
Introducing spherical coordinates in the orthogonal three-space,
\begin{equation}
\begin{split}
dY^2=dR^2+R^2d\Omega_2^2,
\end{split}
\end{equation}
one arrives at
\begin{equation}
\begin{split}
ds^2=-dT^2+dR^2+R^2d\Omega_2^2.
\end{split}
\end{equation}
Thus the adapted chart is nothing but the usual spherical rest frame of the charge, written in a Lorentz-invariant manner.

\subsection{Coulomb field in the adapted frame}
\label{subsec:flat-coulomb-adapted}

In the geodesic-centered frame the physical content of the problem is immediate: the charge is static at $R=0$, so the electromagnetic field is purely Coulombic. We take the gauge potential in the electrostatic form
\begin{equation}
\begin{split}
A=-\phi(R)\,dT.
\end{split}
\end{equation}
Then the field strength is
\begin{equation}
\begin{split}
F=dA=-\phi'(R)\,dR\wedge dT.
\end{split}
\end{equation}
Away from the source, Maxwell's equations reduce to the radial Laplace equation,
\begin{equation}
\begin{split}
\frac{1}{R^2}\frac{d}{dR}\Bigl(R^2\phi'(R)\Bigr)=0.
\end{split}
\end{equation}
Hence
\begin{equation}
\begin{split}
R^2\phi'(R)=\text{constant},
\end{split}
\end{equation}
and matching to the standard point-charge normalization gives
\begin{equation}
\begin{split}
\phi(R)=\frac{q}{4\pi R}.
\end{split}
\end{equation}
Therefore
\begin{equation}
\begin{split}
A=-\frac{q}{4\pi R}\,dT,
\qquad
F=dA=\frac{q}{4\pi R^2}\,dR\wedge dT.
\end{split}
\end{equation}
Equivalently,
\begin{equation}
\begin{split}
F_{RT}=+\frac{q}{4\pi R^2}.
\end{split}
\end{equation}

At this stage the physics is exactly that of a static Coulomb field. All effects associated with the uniform motion of the charge are encoded not in the local form of the solution in the adapted frame, but in how the coordinates $(T,R,\Omega)$ are embedded inside the original Minkowski coordinates.

\subsection{Field strength in inertial coordinates}
\label{subsec:flat-field-inertial}

We now express the Coulomb field in an arbitrary inertial frame. Since
\begin{equation}
\begin{split}
T=-u\cdot x,
\qquad
R^2=x^2+(u\cdot x)^2,
\end{split}
\end{equation}
we have
\begin{equation}
\begin{split}
dT=-u_\mu dx^\mu.
\end{split}
\end{equation}
Differentiating $R^2$ gives
\begin{equation}
\begin{split}
2R\,dR
&=
d\bigl(x^2+(u\cdot x)^2\bigr) \\
&=
2x_\mu dx^\mu+2(u\cdot x)u_\mu dx^\mu,
\end{split}
\end{equation}
and hence
\begin{equation}
\begin{split}
dR
=
\frac{x_\mu+(u\cdot x)u_\mu}{R}\,dx^\mu.
\end{split}
\end{equation}
Substituting into
\begin{equation}
\begin{split}
F=\frac{q}{4\pi R^2}\,dR\wedge dT
\end{split}
\end{equation}
yields
\begin{equation}
\begin{split}
F
&=
\frac{q}{4\pi R^3}
\bigl(x_\mu+(u\cdot x)u_\mu\bigr)(-u_\nu)\,dx^\mu\wedge dx^\nu.
\end{split}
\end{equation}
The term proportional to $u_\mu u_\nu$ drops out by antisymmetry, leaving
\begin{equation}
\begin{split}
F
=
-\frac{q}{4\pi R^3}x_\mu u_\nu\,dx^\mu\wedge dx^\nu.
\end{split}
\end{equation}
Therefore the covariant field strength is
\begin{equation}
\begin{split}
F_{\mu\nu}
=
\frac{q}{4\pi R^3}\bigl(x_\nu u_\mu-x_\mu u_\nu\bigr),
\qquad
R=\sqrt{x^2+(u\cdot x)^2}.
\end{split}
\end{equation}

This is the familiar field of a uniformly moving charge, but it has been obtained here without any explicit Lorentz boost of the static solution. Instead, it follows directly from the statement that uniform motion corresponds to a timelike geodesic and that the appropriate geometric description is a Coulomb field written in coordinates centered on that geodesic.


To connect with the asymptotic matching problem, we now extract the radial electric field in standard spherical coordinates. Let
\begin{equation}
\begin{split}
x^\mu=(t,\vec x),
\qquad
\vec x=r\hat x,
\qquad
\hat x^2=1.
\end{split}
\end{equation}
Using
\begin{equation}
\begin{split}
u^\mu=\gamma(1,\vec\beta),
\qquad
u_\mu=(-\gamma,\gamma\vec\beta),
\end{split}
\end{equation}
the electric components are
\begin{equation}
\begin{split}
F_{i0}
=
\frac{q}{4\pi R^3}\bigl(x_0u_i-x_i u_0\bigr).
\end{split}
\end{equation}
Since
\begin{equation}
\begin{split}
x_0=-t,
\qquad
x_i=r\hat x_i,
\end{split}
\end{equation}
this becomes
\begin{equation}
\begin{split}
F_{i0}
&=
\frac{q}{4\pi R^3}\bigl((-t)\gamma\beta_i-r\hat x_i(-\gamma)\bigr) \\
&=
\frac{q\gamma}{4\pi R^3}\bigl(r\hat x_i-t\beta_i\bigr).
\end{split}
\end{equation}
Projecting onto the radial direction,
\begin{equation}
\begin{split}
F_{rt}:=\hat x^iF_{i0},
\end{split}
\end{equation}
we obtain
\begin{equation}
\begin{split}
F_{rt}
=
\frac{q\gamma}{4\pi R^3}\bigl(r-t\,\hat x\cdot\vec\beta\bigr).
\end{split}
\end{equation}

To make the denominator explicit, note that
\begin{equation}
\begin{split}
x^2=-t^2+r^2,
\end{split}
\end{equation}
while
\begin{equation}
\begin{split}
u\cdot x
&=
-\gamma t+\gamma\vec\beta\cdot\vec x \\
&=
-\gamma\bigl(t-r\,\hat x\cdot\vec\beta\bigr).
\end{split}
\end{equation}
Hence
\begin{equation}
\begin{split}
R^2
=
-t^2+r^2+\gamma^2\bigl(t-r\,\hat x\cdot\vec\beta\bigr)^2.
\end{split}
\end{equation}
Therefore
\begin{equation}
\begin{split}
F_{rt}(t,r,\hat x)
=
\frac{q\gamma\bigl(r-t\,\hat x\cdot\vec\beta\bigr)}
{4\pi\Bigl[\gamma^2\bigl(t-r\,\hat x\cdot\vec\beta\bigr)^2-t^2+r^2\Bigr]^{3/2}}.
\end{split}
\end{equation}

This expression makes clear that even though the field remains Coulombic in origin, its angular distribution in an inertial frame is anisotropic. Physically, the anisotropy is not a radiative effect. Rather, it is the boosted Coulomb profile of the moving charge, and it is precisely this anisotropic Coulombic data that survives at the corners of null infinity and enters the matching problem.

\subsection{Future null infinity and the Coulombic data at \texorpdfstring{$\mathscr I^+_-$}{I+}}
\label{subsec:flat-future-null-infinity}

We now examine the field near future null infinity. Introduce retarded time
\begin{equation}
\begin{split}
u=t-r,
\end{split}
\end{equation}
so that
\begin{equation}
\begin{split}
t=u+r.
\end{split}
\end{equation}
We take the limit $r\to\infty$ at fixed $u$ and fixed angle $\hat x$.

The numerator of $F_{rt}$ behaves as
\begin{equation}
\begin{split}
r-t\,\hat x\cdot\vec\beta
&=
r-(u+r)\hat x\cdot\vec\beta \\
&=
r(1-\hat x\cdot\vec\beta)-u\,\hat x\cdot\vec\beta \\
&\sim
r(1-\hat x\cdot\vec\beta).
\end{split}
\end{equation}
For the denominator we write
\begin{equation}
\begin{split}
t-r\,\hat x\cdot\vec\beta
&=
u+r-r\,\hat x\cdot\vec\beta \\
&=
u+r(1-\hat x\cdot\vec\beta),
\end{split}
\end{equation}
which gives
\begin{equation}
\begin{split}
R^2
&=
\gamma^2\bigl(u+r(1-\hat x\cdot\vec\beta)\bigr)^2-(u+r)^2+r^2.
\end{split}
\end{equation}
At leading order,
\begin{equation}
\begin{split}
R^2\sim r^2\gamma^2(1-\hat x\cdot\vec\beta)^2,
\qquad
R\sim r\,\gamma(1-\hat x\cdot\vec\beta).
\end{split}
\end{equation}
Substituting into the field strength yields
\begin{equation}
\begin{split}
F_{rt}
\sim
\frac{q}{4\pi r^2\gamma^2(1-\hat x\cdot\vec\beta)^2}.
\end{split}
\end{equation}
Since on $\mathscr I^+$ one identifies
\begin{equation}
\begin{split}
F_{ru}=F_{rt},
\end{split}
\end{equation}
it follows that
\begin{equation}
\begin{split}
\lim_{r\to\infty}r^2F_{ru}(u,\hat x)\Big|_{\mathscr I^+}
=
\frac{q}{4\pi\gamma^2(1-\hat x\cdot\vec\beta)^2}.
\end{split}
\end{equation}
This leading coefficient is independent of $u$, and therefore the same expression gives the Coulombic data at the past endpoint of future null infinity
\begin{equation}
\begin{split}
\lim_{r\to\infty}r^2F_{ru}(u,\hat x)\Big|_{\mathscr I^+_-}
=
\frac{q}{4\pi\gamma^2(1-\hat x\cdot\vec\beta)^2}.
\end{split}
\end{equation}

The independence of $u$ is physically important. It signals that we are isolating the nonradiative, deeply infrared Coulombic part of the solution. This is the part that carries the long-range memory of the charge distribution and controls the matching across spatial infinity.

\subsection{Past null infinity and the Coulombic data at \texorpdfstring{$\mathscr I^-_+$}{I-}}
\label{subsec:flat-past-null-infinity}

We next approach past null infinity using advanced time
\begin{equation}
\begin{split}
v=t+r,
\end{split}
\end{equation}
so that
\begin{equation}
\begin{split}
t=v-r.
\end{split}
\end{equation}
Again we take $r\to\infty$ at fixed $v$ and fixed angle $\hat x$.

The numerator behaves as
\begin{equation}
\begin{split}
r-t\,\hat x\cdot\vec\beta
&=
r-(v-r)\hat x\cdot\vec\beta \\
&=
r(1+\hat x\cdot\vec\beta)-v\,\hat x\cdot\vec\beta \\
&\sim
r(1+\hat x\cdot\vec\beta).
\end{split}
\end{equation}
Likewise,
\begin{equation}
\begin{split}
t-r\,\hat x\cdot\vec\beta
&=
v-r-r\,\hat x\cdot\vec\beta \\
&=
v-r(1+\hat x\cdot\vec\beta),
\end{split}
\end{equation}
and hence
\begin{equation}
\begin{split}
R^2
&=
\gamma^2\bigl(v-r(1+\hat x\cdot\vec\beta)\bigr)^2-(v-r)^2+r^2 \\
&\sim
r^2\gamma^2(1+\hat x\cdot\vec\beta)^2.
\end{split}
\end{equation}
Thus
\begin{equation}
\begin{split}
R\sim r\,\gamma(1+\hat x\cdot\vec\beta),
\end{split}
\end{equation}
and so
\begin{equation}
\begin{split}
F_{rt}
\sim
\frac{q}{4\pi r^2\gamma^2(1+\hat x\cdot\vec\beta)^2}.
\end{split}
\end{equation}
On $\mathscr I^-$ one has
\begin{equation}
\begin{split}
F_{rv}=F_{rt},
\end{split}
\end{equation}
which gives
\begin{equation}
\begin{split}
\lim_{r\to\infty}r^2F_{rv}(v,\hat x)\Big|_{\mathscr I^-}
=
\frac{q}{4\pi\gamma^2(1+\hat x\cdot\vec\beta)^2}.
\end{split}
\end{equation}
Again this is independent of $v$, and therefore it is also the field at the future endpoint of past null infinity:
\begin{equation}
\begin{split}
\lim_{r\to\infty}r^2F_{rv}(v,\hat x)\Big|_{\mathscr I^-_+}
=
\frac{q}{4\pi\gamma^2(1+\hat x\cdot\vec\beta)^2}.
\end{split}
\end{equation}

At this stage one sees the key subtlety of the infrared problem. The field values at $\mathscr I^+_-$ and $\mathscr I^-_+$ are both perfectly well defined, but they are not equal when compared at the same angle. In other words, the Coulombic field is not continuous at spatial infinity in the naive sense.

\subsection{Antipodal matching condition}
\label{subsec:flat-antipodal-matching}

We now compare the two asymptotic data sets. At fixed angle $\hat x$,
\begin{equation}
\begin{split}
\lim_{r\to\infty}r^2F_{ru}(u,\hat x)\Big|_{\mathscr I^+_-}
=
\frac{q}{4\pi\gamma^2(1-\hat x\cdot\vec\beta)^2},
\end{split}
\end{equation}
whereas
\begin{equation}
\begin{split}
\lim_{r\to\infty}r^2F_{rv}(v,\hat x)\Big|_{\mathscr I^-_+}
=
\frac{q}{4\pi\gamma^2(1+\hat x\cdot\vec\beta)^2}.
\end{split}
\end{equation}
These expressions differ unless the velocity vanishes or the observation angle is special. Therefore the leading Coulombic field is not continuous across $i^0$ when the same angular coordinate is used on the two celestial spheres.

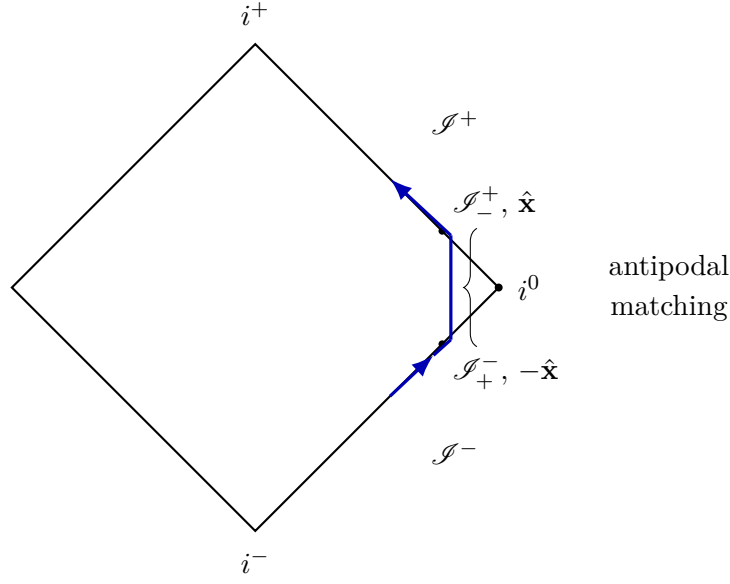
\begin{figure}[H]
\centering
\begin{tikzpicture}[scale=1.15,>=Latex, every node/.style={align=center}]

\draw[thick] (0,2.8) -- (2.8,0) -- (0,-2.8) -- (-2.8,0) -- cycle;

\node at (0,3.15) {$i^+$};
\node at (0,-3.15) {$i^-$};
\node at (3.15,0) {$i^0$};

\node[right] at (1.9,1.9) {$\mathscr I^+$};
\node[right] at (1.9,-1.9) {$\mathscr I^-$};

\fill (2.15,0.65) circle (1.2pt);
\fill (2.15,-0.65) circle (1.2pt);

\node[above right] at (2.15,0.65) {$\mathscr I^+_{-},\,\hat{\mathbf x}$};
\node[below right] at (2.15,-0.65) {$\mathscr I^-_{+},\,-\hat{\mathbf x}$};


\draw[very thick,blue!70!black,->] (1.55,-1.25) -- (2.05,-0.78);
\draw[very thick,blue!70!black] (2.05,-0.78) -- (2.25,-0.60);
\draw[very thick,blue!70!black] (2.25,-0.60) -- (2.25,0.60);
\draw[very thick,blue!70!black,->] (2.25,0.60) -- (1.55,1.25);

\fill (2.8,0) circle (1.3pt);

\draw[decorate,decoration={brace,amplitude=5pt}] (2.55,-0.68) -- (2.55,0.68);
\node[right] at (3.95,0) {antipodal\\matching};

\node at (0,-3.95) {$\displaystyle
\lim_{r\to\infty} r^2 F_{ru}(u,\hat{\mathbf x})\Big|_{\mathscr I^+_-}
=
\lim_{r\to\infty} r^2 F_{rv}(v,-\hat{\mathbf x})\Big|_{\mathscr I^-_+}
$};

\end{tikzpicture}
\caption{Penrose diagram illustrating the antipodal matching of the leading Coulombic part of the Li\'enard--Wiechert field. The field at angle \(\hat{\mathbf x}\) on the past of future null infinity \(\mathscr I^+_-\) matches the field at the antipodal angle \(-\hat{\mathbf x}\) on the future of past null infinity \(\mathscr I^-_+\).}
\label{fig:antipodal_matching_flat}
\end{figure}
The correct matching is instead antipodal. Under the antipodal map
\begin{equation}
\begin{split}
\hat x\mapsto -\hat x,
\nonumber
\end{split}
\end{equation}
one has
\begin{equation}
\begin{split}
(-\hat x)\cdot\vec\beta=-\hat x\cdot\vec\beta,
\end{split}
\nonumber
\end{equation}
and hence
\begin{equation}
\begin{split}
\lim_{r\to\infty}r^2F_{rv}(v,-\hat x)\Big|_{\mathscr I^-_+}
=
\frac{q}{4\pi\gamma^2(1-\hat x\cdot\vec\beta)^2}.
\end{split}
\end{equation}
This is precisely the value found at $\mathscr I^+_-$. We therefore obtain the antipodal matching condition
\begin{equation}
\begin{split}
\lim_{r\to\infty}r^2F_{ru}(u,\hat x)\Big|_{\mathscr I^+_-}
=
\lim_{r\to\infty}r^2F_{rv}(v,-\hat x)\Big|_{\mathscr I^-_+}
.
\end{split}
\end{equation}

This is the flat-space matching law for the leading Coulombic field strength. See Figure \ref{fig:antipodal_matching_flat}. 

The geodesic-centered derivation highlights the physical origin of antipodal matching in a particularly transparent way. A uniformly moving charge is still, in an intrinsic sense, a Coulombic source. The apparent complexity of the Liénard--Wiechert field is largely a consequence of describing this Coulomb field in a frame not centered on the source geodesic. Once the solution is first written in the adapted frame, the angular dependence at null infinity arises entirely from the embedding of that rest-frame Coulomb profile into an arbitrary inertial slicing. From this viewpoint, the factors
\begin{equation}
\begin{split}
\frac{1}{\gamma^2(1\mp \hat x\cdot\vec\beta)^2}
\end{split}
\end{equation}
have a direct geometric interpretation: they are the asymptotic imprint of the rest-frame radial field on the future and past celestial spheres. The mismatch at fixed angle and the restoration of equality after the antipodal map are then two aspects of the same phenomenon. The Coulombic field is anisotropic in inertial coordinates, yet remains continuous along the null generators of conformal infinity. This is precisely the content of the antipodal matching condition and the reason it is the correct infrared boundary condition for the electromagnetic field in flat-spacetime. 

\subsection{Conformal compactification, image charges, and antipodal matching}
\label{subsec:flat-conformal-image}
A useful geometric way to understand the antipodal matching condition is to use the
conformal invariance of four-dimensional electromagnetism. Since Maxwell theory in four
dimensions is invariant under Weyl rescalings of the metric, one may conformally compactify
Minkowski space into a finite diamond embedded in the Einstein static universe
\(S^{3}\times \mathbb{R}\). In this compactified description, the cross-sections of the
Einstein cylinder are \(S^{3}\), and the physical Minkowski spacetime occupies a diamond-shaped
region whose boundary is null infinity. Spatial infinity \(i^{0}\) sits at the point where the
past and future null boundaries meet. More precisely, \(\mathscr I\) is the lightcone of
\(i^{0}\) in the compactified geometry: the past lightcone of \(i^{0}\) gives
\(\mathscr I^{-}\), while the future lightcone of \(i^{0}\) gives \(\mathscr I^{+}\). A null
generator therefore begins on \(\mathscr I^{-}\), passes through \(i^{0}\), and continues to
\(\mathscr I^{+}\). Because the generator emerges on the future celestial sphere at the
antipodal angular point, the natural Lorentz-invariant gluing condition is not equality at
the same angle, but equality at antipodal angles. Thus the leading Coulombic field satisfies
\begin{equation}
\label{eq:antipodal_matching_condition}
\left.
\lim_{r\to\infty} r^{2}F_{ru}(u,\hat x)
\right|_{\mathscr I^{+}_{-}}
=
\left.
\lim_{r\to\infty} r^{2}F_{rv}(v,-\hat x)
\right|_{\mathscr I^{-}_{+}} .
\end{equation}
This is the antipodal matching condition. It states that the radial electric field is continuous
along the null generators of the conformal boundary as they pass through spatial infinity,
rather than continuous at fixed angular coordinate. This distinction is essential: matching
at the same angle would not be Lorentz invariant, while the antipodal matching condition is
compatible with the action of Lorentz transformations on the compactified geometry
\cite{Strominger:2017zoo}.

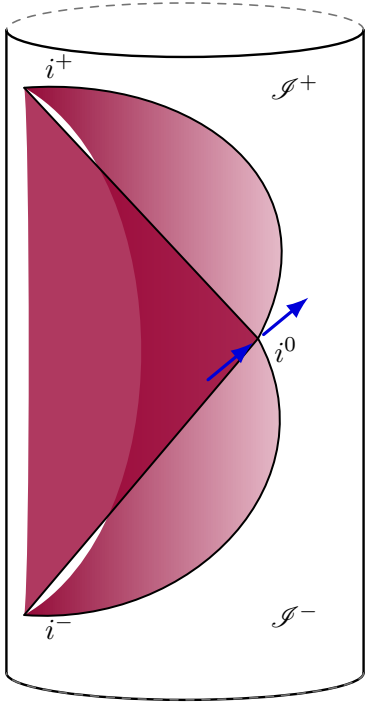
\begin{figure}[H]
\centering
\begin{tikzpicture}[
    scale=1.05,
    line cap=round,
    line join=round,
    >=Latex
]

\def\R{2.25}    
\def\H{8.10}    
\def\e{0.34}    

\coordinate (TopL) at (-0.90*\R,0.91*\H);
\coordinate (BotL) at (-0.90*\R,0.09*\H);
\coordinate (Mid)  at (0.92,0.52*\H);

\begin{scope}
\clip
(-\R,0) --
(-\R,\H)
arc[start angle=180,end angle=0,x radius=\R,y radius=\e] --
(\R,0)
arc[start angle=0,end angle=-180,x radius=\R,y radius=\e] --
cycle;

\shade[
    left color=purple!75!black,
    right color=purple!12,
    opacity=0.97
]
(TopL)
.. controls (0.15,0.93*\H) and (1.95,0.76*\H) .. (Mid)
.. controls (1.90,0.30*\H) and (0.10,0.07*\H) .. (BotL)
.. controls (-0.06,0.23*\H) and (-0.06,0.77*\H) .. cycle;

\fill[purple!80!black,opacity=0.78]
(TopL) -- (Mid) -- (BotL)
.. controls (-1.95,0.16*\H) and (-1.95,0.84*\H) .. cycle;

\draw[line width=0.75pt]
(TopL)
.. controls (0.15,0.93*\H) and (1.95,0.76*\H) .. (Mid)
.. controls (1.90,0.30*\H) and (0.10,0.07*\H) .. (BotL);

\draw[line width=0.75pt] (TopL) -- (Mid) -- (BotL);

\end{scope}

\draw[line width=0.9pt] (-\R,0) -- (-\R,\H);
\draw[line width=0.9pt] (\R,0) -- (\R,\H);

\draw[line width=0.9pt] (-\R,\H) arc[start angle=180,end angle=360,x radius=\R,y radius=\e];
\draw[dashed, black!55, line width=0.6pt]
(\R,\H) arc[start angle=0,end angle=180,x radius=\R,y radius=\e];

\draw[line width=0.9pt] (-\R,0) arc[start angle=180,end angle=360,x radius=\R,y radius=\e];
\draw[dashed, black!55, line width=0.6pt]
(\R,0) arc[start angle=0,end angle=-180,x radius=\R,y radius=\e];

\draw[blue!85!black, very thick, -{Latex[length=3mm,width=2.2mm]}]
($(Mid)+(-0.63,-0.52)$) -- ($(Mid)+(-0.06,-0.05)$);

\draw[blue!85!black, very thick, -{Latex[length=3mm,width=2.2mm]}]
($(Mid)+(0.07,0.05)$) -- ($(Mid)+(0.64,0.52)$);

\node at (-1.58,7.65) {$i^{+}$};
\node at (-1.58,0.58) {$i^{-}$};
\node[anchor=west] at ($(Mid)+(0.08,-0.14)$) {$i^{0}$};

\node at (1.38,7.38) {$\mathscr{I}^{+}$};
\node at (1.38,0.72) {$\mathscr{I}^{-}$};

\end{tikzpicture}
\caption{
The purple region is Minkowski spacetime conformally compactified into the Einstein static universe \(S^3\times \mathbb{R}\), represented schematically as a cylinder. The future and past null boundaries are denoted by \(\mathscr I^\pm\), while \(i^\pm\) and \(i^0\) are future timelike, past timelike, and spatial infinity, respectively. The blue arrows indicate that null generators pass continuously through \(i^0\), which underlies the antipodal matching condition.
}
\label{fig:compactified_minkowski_cylinder_purple}
\end{figure}
There is an important subtlety at \(i^{0}\). Although the antipodal matching condition imposes
continuity of the leading field along null generators, it does not require the Maxwell field to
be smooth at spatial infinity. In fact, \(i^{0}\) is generically singular. The reason can be seen
already for a single electric charge in the interior of Minkowski space. After conformal
compactification, a spatial slice becomes a compact \(S^{3}\). On a compact space without
boundary, Gauss's law forbids an isolated net charge unless there is a compensating source
somewhere else on the compact slice. In the compactified Minkowski geometry, this
compensating source is naturally located at \(i^{0}\). Thus a single charge in the physical
interior is accompanied, in the compactified description, by an image charge at spatial
infinity. See Appendix \ref{imch} for more deatils. The image charge is not an additional physical charge in the original Minkowski
spacetime; it is the way the long-range Coulomb field is represented after compactification.
Its presence explains why the electric field generically diverges or becomes singular near
\(i^{0}\). See Figure \ref{fig:compactified_minkowski_cylinder_purple}.

For several moving charges, the singularity at spatial infinity can be much more
complicated, because the angular profile of the Coulombic field depends on the velocities
and charges of all the particles. Therefore one cannot demand ordinary smoothness at
\(i^{0}\). The correct condition is weaker and more physical: the leading Coulombic data must
be transported continuously along the null generators through \(i^{0}\), which is precisely the
antipodal matching condition. This is the geometric
origin of the matching rule used in the infrared analysis of gauge theory and gravity
\cite{Strominger:2017zoo}.

\section{Global \texorpdfstring{$\mathrm{AdS}_4$}{AdS4} and the distinguished null-fringe times}
\label{sec:ads-null-fringe-times}

We begin with Lorentzian global \(\mathrm{AdS}_4\), realized as the hyperboloid
\begin{equation}
\begin{split}
-(X^1)^2-(X^2)^2+(X^3)^2+(X^4)^2+(X^5)^2=-L^2
\end{split}
\end{equation}
embedded in \(\mathbb{R}^{3,2}\) with metric
\begin{equation}
\begin{split}
\eta_{AB}=\mathrm{diag}(-1,-1,+1,+1,+1).
\end{split}
\end{equation}
A convenient global parametrization is
\begin{equation}
\begin{split}
X^1&=\frac{L\cos\tau}{\cos\rho},
\qquad
X^2=\frac{L\sin\tau}{\cos\rho}, \\
X^3&=L\tan\rho\,\sin\theta\cos\phi, \\
X^4&=L\tan\rho\,\sin\theta\sin\phi, \\
X^5&=L\tan\rho\,\cos\theta,
\end{split}
\end{equation}
for which the induced metric takes the standard form
\begin{equation}
\begin{split}
ds^2
=
\frac{L^2}{\cos^2\rho}
\Bigl(
-d\tau^2+d\rho^2+\sin^2\rho\,d\Omega_2^2
\Bigr),
\qquad
0\le \rho<\frac{\pi}{2}.
\end{split}
\end{equation}
The conformal boundary lies at
\begin{equation}
\begin{split}
\rho=\frac{\pi}{2},
\end{split}
\end{equation}
and after stripping off the divergent Weyl factor \(L^2/\cos^2\rho\), one obtains the boundary cylinder
\begin{equation}
\begin{split}
ds^2_{\partial\mathrm{AdS}}
=
-d\tau^2+d\Omega_2^2.
\end{split}
\end{equation}

The times
\begin{equation}
\begin{split}
\tau=\pm \frac{\pi}{2}
\end{split}
\end{equation}
are distinguished by the causal structure of global AdS. Indeed, for radial null motion one sets
\begin{equation}
\begin{split}
d\Omega_2^2=0,
\qquad
ds^2=0,
\end{split}
\end{equation}
so that the metric implies
\begin{equation}
\begin{split}
-d\tau^2+d\rho^2=0
\qquad\Longrightarrow\qquad
d\tau=\pm d\rho.
\end{split}
\end{equation}
A radial null ray emitted from the center \(\rho=0\) at \(\tau=0\) therefore reaches the boundary \(\rho=\pi/2\) at
\begin{equation}
\begin{split}
\tau=\pm \frac{\pi}{2}.
\end{split}
\end{equation}

This simple observation is central for the entire construction. The neighborhoods of the boundary cylinder near
\begin{equation}
\begin{split}
\tau=+\frac{\pi}{2}
\qquad\text{and}\qquad
\tau=-\frac{\pi}{2}
\end{split}
\end{equation}
are the regions singled out by radial null propagation from the bulk center. In the large-radius limit \(L\to\infty\), these thin strips are precisely the regions that become the flat-space avatars of future and past null infinity. For this reason, they play the role of future and past ``null fringes'' already at finite AdS radius.

This geometric fact is the origin of the later matching problem. Since the electromagnetic field sourced by a bulk charge reaches the boundary along null directions, the leading Coulombic data naturally organize themselves near these two distinguished boundary times. The task is then to compute those data and compare the future and past limits.

\section{Maxwell theory on a fixed global \texorpdfstring{$\mathrm{AdS}_4$}{AdS4} background}
\label{sec:maxwell-global-ads}

We now turn to Maxwell theory on a fixed global \(\mathrm{AdS}_4\) background. The action is
\begin{equation}
S_{\mathrm{Maxwell}}
=
-\frac14
\int d^4x\,\sqrt{-g}\,F_{\mu\nu}F^{\mu\nu}
+
\int d^4x\,\sqrt{-g}\,A_\mu J^\mu,
\qquad
F=dA.
\end{equation}
Varying the gauge potential gives the equations of motion
\begin{equation}
\nabla_\mu F^{\mu\nu}=J^\nu.
\end{equation}
We are interested in the field of a point charge localized at the center of AdS. Away from the source, the current vanishes, and the field therefore satisfies the homogeneous Maxwell equation
\begin{equation}
\nabla_\mu F^{\mu\nu}=0.
\end{equation}

Since \(F=dA\), the field strength is determined by ordinary derivatives,
\begin{equation}
F_{\mu\nu}=\partial_\mu A_\nu-\partial_\nu A_\mu.
\end{equation}
Equivalently,
\begin{equation}
F_{\mu\nu}
=
\nabla_\mu A_\nu-\nabla_\nu A_\mu,
\end{equation}
because the Levi-Civita connection is torsion free and the Christoffel symbols cancel under antisymmetrization. 

For later use, let us record the basic metric data in global AdS. Starting from
\begin{equation}
ds^2
=
\frac{L^2}{\cos^2\rho}
\Bigl(
-d\tau^2+d\rho^2+\sin^2\rho\,d\Omega_2^2
\Bigr),
\end{equation}
one reads off
\begin{equation}
g_{\tau\tau}=-\frac{L^2}{\cos^2\rho},
\qquad
g_{\rho\rho}=\frac{L^2}{\cos^2\rho},
\qquad
g_{\theta\theta}=\frac{L^2}{\cos^2\rho}\sin^2\rho,
\qquad
g_{\phi\phi}=\frac{L^2}{\cos^2\rho}\sin^2\rho\,\sin^2\theta,
\end{equation}
and hence
\begin{equation}
g^{\tau\tau}=-\frac{\cos^2\rho}{L^2},
\qquad
g^{\rho\rho}=\frac{\cos^2\rho}{L^2},
\qquad
g^{\theta\theta}=\frac{\cos^2\rho}{L^2\sin^2\rho},
\qquad
g^{\phi\phi}=\frac{\cos^2\rho}{L^2\sin^2\rho\,\sin^2\theta}.
\end{equation}
The volume factor is
\begin{equation}
\sqrt{-g}
=
\frac{L^4}{\cos^4\rho}\sin^2\rho\,\sin\theta.
\end{equation}
Therefore Maxwell's equation may be written in divergence form as
\begin{equation}
\nabla_\mu F^{\mu\nu}
=
\frac{1}{\sqrt{-g}}
\partial_\mu\!\Bigl(\sqrt{-g}\,F^{\mu\nu}\Bigr).
\end{equation}

The electrostatic solution for a charge held at the center of AdS will serve as the seed for the more general moving-charge solution. The crucial conceptual point is that the moving problem does not require solving Maxwell's equation again from scratch. Rather, it is obtained by recentering this static solution on an arbitrary timelike geodesic.

\subsection{Static Coulomb field at the center of global \texorpdfstring{$\mathrm{AdS}_4$}{AdS4}}
\label{subsec:ads-static-coulomb-center}

We now solve explicitly for the electrostatic field of a point charge sitting at the center \(\rho=0\). Since the source is at rest and located at the center of the rotationally invariant geometry, the solution must be static and spherically symmetric. The most general ansatz compatible with these symmetries is
\begin{equation}
A=A_\tau(\rho)\,d\tau,
\qquad
A_\rho=A_\theta=A_\phi=0.
\end{equation}
The corresponding field strength is
\begin{equation}
F=dA
=
(\partial_\rho A_\tau)\,d\rho\wedge d\tau,
\qquad
F_{\rho\tau}=\partial_\rho A_\tau.
\end{equation}
Only the \(\tau\)-component of Maxwell's equation can be nontrivial, so the field equation reduces to
\begin{equation}
\nabla_\mu F^{\mu\tau}=0.
\end{equation}
Using the divergence form and spherical symmetry,
\begin{equation}
\frac{1}{\sqrt{-g}}
\partial_\rho\!\Bigl(\sqrt{-g}\,F^{\rho\tau}\Bigr)=0.
\end{equation}
Now
\begin{equation}
F^{\rho\tau}
=
g^{\rho\rho}g^{\tau\tau}F_{\rho\tau}
=
-\frac{\cos^4\rho}{L^4}\,\partial_\rho A_\tau.
\end{equation}
Multiplying by \(\sqrt{-g}\), one finds
\begin{equation}
\sqrt{-g}\,F^{\rho\tau}
=
\frac{L^4}{\cos^4\rho}\sin^2\rho\,\sin\theta
\left(
-\frac{\cos^4\rho}{L^4}\,\partial_\rho A_\tau
\right)
=
-\sin^2\rho\,\sin\theta\,\partial_\rho A_\tau.
\end{equation}
The angular factor \(\sin\theta\) is a spectator, so the field equation becomes the one-dimensional radial equation
\begin{equation}
\partial_\rho\!\Bigl(\sin^2\rho\,\partial_\rho A_\tau\Bigr)=0.
\end{equation}

This integrates immediately:
\begin{equation}
\sin^2\rho\,\partial_\rho A_\tau=C,
\qquad
\partial_\rho A_\tau=C\,\csc^2\rho,
\end{equation}
and hence
\begin{equation}
A_\tau(\rho)=a+b\cot\rho,
\end{equation}
where \(a\) and \(b\) are constants. The additive constant \(a\) is pure gauge and may be dropped.

The remaining coefficient is fixed by matching to the local flat-space Coulomb solution near the AdS center. For \(\rho\ll1\), one has
\begin{equation}
r=L\tan\rho\sim L\rho,
\qquad
t=L\tau,
\qquad
A_t=\frac{A_\tau}{L}.
\end{equation}
We normalize the flat-space rest-frame Coulomb potential as
\begin{equation}
\phi(r)=\frac{q}{4\pi r},
\qquad
A=-\phi(r)\,dt.
\end{equation}
Therefore
\begin{equation}
A_t=-\frac{q}{4\pi r},
\end{equation}
and near \(\rho=0\),
\begin{equation}
A_\tau
=
L A_t
\sim
-\frac{Lq}{4\pi r}
\sim
-\frac{q}{4\pi\rho}.
\end{equation}
Since
\begin{equation}
\cot\rho\sim \frac1\rho
\qquad
(\rho\to0),
\end{equation}
it follows that
\begin{equation}
b=-\frac{q}{4\pi}.
\end{equation}
Hence the static Coulomb potential in global AdS is
\begin{equation}
A
=
-\frac{q}{4\pi}\cot\rho\,d\tau.
\end{equation}

Its field strength is
\begin{equation}
F=dA
=
\frac{q}{4\pi}\csc^2\rho\,d\rho\wedge d\tau,
\end{equation}
so the nonzero covariant components are
\begin{equation}
F_{\rho\tau}=+\frac{q}{4\pi}\csc^2\rho,
\qquad
F_{\tau\rho}=-\frac{q}{4\pi}\csc^2\rho.
\end{equation}

This solution has the expected properties. Near \(\rho=0\), it reproduces the flat-space Coulomb singularity with the same normalization convention as the Minkowski computation. Away from the source, it solves the homogeneous Maxwell equation, so the electric flux through any sphere of fixed \(\rho\) is conserved. Most importantly, it is the unique seed from which the field of a charge moving on an arbitrary timelike AdS geodesic will be obtained.

\section{Timelike geodesics and geodesic-adapted coordinates in \texorpdfstring{$\mathrm{AdS}_4$}{AdS4}}
\label{sec:ads-geodesic-coordinates}

The central geometric idea of the moving-charge construction is that in AdS space the natural analogue of uniform inertial motion in Minkowski space is motion along a timelike geodesic. In flat-space, the electromagnetic field of a uniformly moving charge may be understood as the ordinary Coulomb field written in a Lorentz frame adapted to the source worldline. The AdS generalization is that the field of a freely moving charge is obtained by expressing the static Coulomb field in coordinates centered on the corresponding timelike geodesic.

This viewpoint shifts the problem from one of repeatedly solving Maxwell's equation for different source trajectories to one of geometry: solve the Coulomb problem once at the center of global AdS, then use the symmetries of AdS to reinterpret that solution in a frame adapted to an arbitrary geodesic. The technical heart of the construction is therefore not a new solution of the field equations, but the invariant reconstruction of the geodesic-adapted coordinates in terms of the original global coordinates.

\subsection{Timelike geodesics from the embedding-space point of view}
\label{subsec:ads-geodesics-embedding}

A particularly efficient way to describe timelike geodesics in \(\mathrm{AdS}_4\) is to work in the ambient space \(\mathbb{R}^{3,2}\), endowed with metric
\begin{equation}
\eta_{AB}=\mathrm{diag}(-1,-1,+1,+1,+1).
\end{equation}
Lorentzian \(\mathrm{AdS}_4\) of radius \(L\) is realized as the hyperboloid
\begin{equation}
X^2\equiv \eta_{AB}X^A X^B=-L^2.
\end{equation}
In this embedding-space description, AdS isometries become ordinary linear transformations preserving \(\eta_{AB}\), and geodesics acquire a simple geometric characterization.

Let \(P\) be a point on the AdS hyperboloid and let \(U\) be a unit timelike tangent vector at \(P\). These satisfy
\begin{equation}
P^2=-L^2,
\qquad
U^2=-1,
\qquad
P\cdot U=0.
\end{equation}
The orthogonality condition \(P\cdot U=0\) is simply the statement that \(U\) lies in the tangent space to the hyperboloid at \(P\). Indeed, the AdS hypersurface is defined by the constraint \(X^2+L^2=0\), so any tangent vector \(V\) at \(P\) must satisfy
\begin{equation}
\left.\frac{d}{d\lambda}\right|_{\lambda=0}(P+\lambda V)^2
=
2P\cdot V=0.
\end{equation}

Given such a pair \((P,U)\), consider the curve
\begin{equation}
Y(s)=P\cos\frac{s}{L}+LU\sin\frac{s}{L}.
\end{equation}
This is the unique timelike geodesic passing through \(P\) with initial tangent \(U\). First,
\begin{equation}
Y(s)^2
=
P^2\cos^2\frac{s}{L}
+2L(P\cdot U)\cos\frac{s}{L}\sin\frac{s}{L}
+L^2U^2\sin^2\frac{s}{L}
=
-L^2,
\end{equation}
so the curve remains on the AdS hyperboloid. Second,
\begin{equation}
Y(0)=P,
\qquad
\dot{Y}(0)=U.
\end{equation}
Third,
\begin{equation}
\dot{Y}(s)
=
-\frac{1}{L}P\sin\frac{s}{L}+U\cos\frac{s}{L},
\end{equation}
and therefore
\begin{equation}
\dot{Y}(s)^2
=
\frac{1}{L^2}P^2\sin^2\frac{s}{L}
-\frac{2}{L}(P\cdot U)\sin\frac{s}{L}\cos\frac{s}{L}
+U^2\cos^2\frac{s}{L}
=
-1.
\end{equation}
Thus \(s\) is proper time. Differentiating once more gives
\begin{equation}
\ddot{Y}(s)=-\frac{1}{L^2}Y(s).
\end{equation}
Since \(Y(s)\) is normal to the hyperboloid, the acceleration is purely normal to AdS and the intrinsic covariant acceleration vanishes. Hence \(Y(s)\) is indeed an intrinsic timelike geodesic.

Geometrically, the curve \(Y(s)\) is the intersection of the AdS hyperboloid with the two-dimensional timelike plane through the origin spanned by \(P\) and \(U\).

\subsection{Physical interpretation: geodesic motion as uniform motion in AdS}
\label{subsec:ads-geodesic-uniform-motion}

In Minkowski space, a freely moving massive particle follows a timelike straight line, and the field of a uniformly moving charge is obtained from the static Coulomb field by a Lorentz transformation. In AdS there is no preferred global notion of constant velocity, but geodesic motion survives as the natural notion of inertial motion. A freely propagating massive source therefore traces a timelike geodesic, and the AdS counterpart of the field of a uniformly moving charge is the field of a charge carried along such a geodesic.

This is the guiding principle behind the later field-theoretic construction. One first solves the static Coulomb problem for a charge at the center of global AdS. One then regards the field of a geodesically moving charge not as a new electrostatic configuration, but as the same Coulomb field viewed in coordinates adapted to the source geodesic.

\subsection{Geodesic-adapted coordinates}
\label{subsec:ads-geodesic-adapted-coordinates}

Let \(\Gamma\) be an arbitrary timelike geodesic. Because \(\mathrm{AdS}_4\) is maximally symmetric, its isometry group acts transitively on unit timelike tangent data. Any geodesic can therefore be mapped by an AdS isometry to the standard central geodesic. Pulling back the usual global coordinates by such an isometry defines a new chart
\begin{equation}
(T,R,\Theta,\Phi),
\end{equation}
which we call the geodesic-adapted coordinates associated with \(\Gamma\).

By construction, in this adapted chart the geodesic \(\Gamma\) sits at the spatial origin,
\begin{equation}
\Gamma:
\qquad
R=0,
\qquad
\Theta,\Phi=\text{constant},
\end{equation}
and, because the coordinates are obtained from the standard global chart by an isometry, the metric retains the same functional form,
\begin{equation}
ds^2
=
\frac{L^2}{\cos^2R}
\Bigl(
-dT^2+dR^2+\sin^2R\,d\Omega_2^2
\Bigr).
\end{equation}
Thus the adapted frame is distinguished: in it the source worldline is literally the central static observer.

This has an immediate physical consequence. In the \((T,R,\Theta,\Phi)\) chart, a point charge moving along \(\Gamma\) is at rest at \(R=0\). Its electromagnetic field must therefore be static and spherically symmetric in these coordinates. Consequently, the gauge potential and field strength are simply the static Coulomb solution already obtained at the AdS center,
\begin{equation}
A
=
-\frac{q}{4\pi}\cot R\,dT,
\qquad
F=dA
=
\frac{q}{4\pi}\csc^2R\,dR\wedge dT.
\end{equation}
Equivalently,
\begin{equation}
F_{RT}=+\frac{q}{4\pi}\csc^2R,
\qquad
F_{TR}=-\frac{q}{4\pi}\csc^2R.
\end{equation}
This is the AdS analogue of the statement that the field of a uniformly moving charge is just the Coulomb field in the rest frame. The nontrivial task is therefore not to solve Maxwell's equation again, but to express the adapted coordinates \((T,R)\) in terms of the original global coordinates \((\tau,\rho,\theta,\phi)\).

\subsection{Invariant reconstruction of the adapted coordinates}
\label{subsec:ads-invariant-reconstruction}

Let \(X\in\mathrm{AdS}_4\) be an arbitrary bulk point, regarded as a vector in embedding space. The geodesic \(\Gamma\) is associated with the timelike two-plane spanned by \(P\) and \(U\), so we decompose \(X\) into components parallel and orthogonal to this plane:
\begin{equation}
X=X_{\parallel}+X_{\perp},
\qquad
X_{\parallel}=\alpha P+\beta U,
\qquad
P\cdot X_{\perp}=U\cdot X_{\perp}=0.
\end{equation}
Taking inner products with \(P\) and \(U\) gives
\begin{equation}
P\cdot X=-L^2\alpha,
\qquad
U\cdot X=-\beta,
\end{equation}
so that
\begin{equation}
\alpha=-\frac{P\cdot X}{L^2},
\qquad
\beta=-(U\cdot X),
\end{equation}
and hence
\begin{equation}
X_{\parallel}
=
-\frac{P\cdot X}{L^2}P-(U\cdot X)U.
\end{equation}

Now introduce the orthonormal timelike basis
\begin{equation}
e_P=\frac{P}{L},
\qquad
e_U=U,
\end{equation}
which satisfies
\begin{equation}
e_P^2=e_U^2=-1,
\qquad
e_P\cdot e_U=0.
\end{equation}
In the geodesic-adapted frame, the same projected vector must also take the form
\begin{equation}
X_{\parallel}
=
\frac{L}{\cos R}
\left(
\cos T\,e_P+\sin T\,e_U
\right).
\end{equation}
Substituting \(e_P=P/L\) and \(e_U=U\), this becomes
\begin{equation}
X_{\parallel}
=
\frac{\cos T}{\cos R}P+\frac{L\sin T}{\cos R}U.
\end{equation}
Comparing coefficients, we obtain
\begin{equation}
-\frac{P\cdot X}{L^2}
=
\frac{\cos T}{\cos R},
\qquad
-(U\cdot X)
=
\frac{L\sin T}{\cos R}.
\end{equation}
Dividing the second by the first yields the adapted time coordinate,
\begin{equation}
\tan T=\frac{L\,U\cdot X}{P\cdot X},
\end{equation}
while squaring and adding gives the adapted radial coordinate,
\begin{equation}
\cos R
=
\frac{L^2}{\sqrt{(P\cdot X)^2+L^2(U\cdot X)^2}}.
\end{equation}
Thus
\begin{equation}
\tan T=\frac{L\,U\cdot X}{P\cdot X},
\qquad
\cos R
=
\frac{L^2}{\sqrt{(P\cdot X)^2+L^2(U\cdot X)^2}}.
\end{equation}

These formulas reconstruct the adapted coordinates entirely in terms of the embedding-space invariants \(P\cdot X\) and \(U\cdot X\). Once \(T\) and \(R\) are known, the moving-charge field follows immediately from the geodesic-centered Coulomb seed,
\begin{equation}
A
=
-\frac{q}{4\pi}\cot R\,dT,
\qquad
F
=
\frac{q}{4\pi}\csc^2R\,dR\wedge dT.
\end{equation}


It is useful to verify that the reconstruction formulas behave correctly on the geodesic itself. Setting
\begin{equation}
X=Y(s)=P\cos\frac{s}{L}+LU\sin\frac{s}{L},
\end{equation}
we find
\begin{equation}
P\cdot Y(s)=-L^2\cos\frac{s}{L},
\qquad
U\cdot Y(s)=-L\sin\frac{s}{L}.
\end{equation}
The invariant formulas then give
\begin{equation}
\tan T=\tan\frac{s}{L},
\qquad
\cos R=1.
\end{equation}
With the continuous branch choice this implies
\begin{equation}
T=\frac{s}{L},
\qquad
R=0,
\end{equation}
exactly as required.

\subsection{Explicit geodesic through the AdS origin}
\label{subsec:ads-explicit-origin-geodesic}

As a concrete example, consider the geodesic passing through the AdS origin at \(\tau=0\) with spatial direction \(\hat p\) and speed \(\beta\). A convenient choice of embedding-space data is
\begin{equation}
P=(L,0,\mathbf{0}),
\qquad
U=(0,\gamma,\gamma\beta\,\hat p),
\qquad
\gamma=\frac{1}{\sqrt{1-\beta^2}}.
\end{equation}
A general bulk point in global AdS coordinates is
\begin{equation}
X
=
\left(
\frac{L\cos\tau}{\cos\rho},
\frac{L\sin\tau}{\cos\rho},
L\tan\rho\,\hat x
\right),
\qquad
\hat x\in S^2.
\end{equation}
The relevant invariants are
\begin{equation}
P\cdot X=-\frac{L^2\cos\tau}{\cos\rho},
\end{equation}
and
\begin{equation}
U\cdot X
=
-\gamma\,\frac{L\sin\tau}{\cos\rho}
+\gamma\beta L\tan\rho\,(\hat p\cdot\hat x)
=
\frac{L\gamma}{\cos\rho}
\Bigl(
-\sin\tau+\beta\sin\rho\,\hat p\cdot\hat x
\Bigr).
\end{equation}
Substituting these into the reconstruction formulas yields
\begin{equation}
\tan T
=
\frac{\gamma\bigl(\sin\tau-\beta\sin\rho\,\hat p\cdot\hat x\bigr)}{\cos\tau},
\end{equation}
and
\begin{equation}
\cos R
=
\frac{\cos\rho}{
\sqrt{
\cos^2\tau+\gamma^2\bigl(\sin\tau-\beta\sin\rho\,\hat p\cdot\hat x\bigr)^2
}
}.
\end{equation}
Therefore the field of a charge moving on this timelike geodesic is obtained simply by substituting these expressions into
\begin{equation}
A
=
-\frac{q}{4\pi}\cot R\,dT,
\qquad
F
=
\frac{q}{4\pi}\csc^2R\,dR\wedge dT.
\end{equation}
In this way the AdS analogue of the Li\'enard--Wiechert field is obtained entirely from the static Coulomb seed by a geometrically natural recentering procedure.

\section{Field strength in global coordinates and antipodal matching in AdS}
\label{sec:ads-field-strength-antipodal}

We now compute the gauge-invariant field strength in the original global coordinates
\((\tau,\rho,\theta,\phi)\), starting from the geodesic-centered Coulomb seed. We work
throughout with the flat-space normalization
\begin{equation}
A_t=-\frac{q}{4\pi r},
\end{equation}
so that in the geodesic-adapted frame \((T,R,\Theta,\Phi)\) the potential and field strength are
\begin{equation}
A=-\frac{q}{4\pi}\cot R\,dT,
\qquad
F=dA=\frac{q}{4\pi}\csc^2R\,dR\wedge dT.
\end{equation}
Equivalently,
\begin{equation}
F_{RT}=+\frac{q}{4\pi}\csc^2R,
\qquad
F_{TR}=-\frac{q}{4\pi}\csc^2R.
\end{equation}

We consider the timelike geodesic through the AdS origin at \(\tau=0\), with spatial
direction \(\hat p\) and speed \(\beta\), and define
\begin{equation}
\gamma=\frac{1}{\sqrt{1-\beta^2}},
\qquad
\mu:=\hat p\cdot \hat x,
\qquad
\vec\beta\cdot\hat x=\beta \mu.
\end{equation}
The geodesic-adapted coordinates are reconstructed in terms of the global coordinates by
\begin{equation}
\tan T
=
\frac{\gamma\bigl(\sin\tau-\beta\sin\rho\,\mu\bigr)}{\cos\tau},
\qquad
\cos R
=
\frac{\cos\rho}{
\sqrt{
\cos^2\tau+\gamma^2\bigl(\sin\tau-\beta\sin\rho\,\mu\bigr)^2
}
}.
\end{equation}

For convenience, let us introduce the shorthand
\begin{equation}
\Xi(\tau,\rho,\hat x):=\sin\tau-\beta\sin\rho\,\mu,
\qquad
\Delta(\tau,\rho,\hat x):=\cos^2\tau+\gamma^2\Xi^2.
\end{equation}
Then
\begin{equation}
\tan T=\frac{\gamma\Xi}{\cos\tau},
\qquad
\cos R=\frac{\cos\rho}{\sqrt{\Delta}}.
\end{equation}
It is also useful to define
\begin{equation}
H(\tau,\rho,\hat x):=\Delta-\cos^2\rho
=
\cos^2\tau-\cos^2\rho+\gamma^2\Xi^2.
\end{equation}
Since
\begin{equation}
\cot R=\frac{\cos R}{\sin R}
=
\frac{\cos\rho}{\sqrt{H}},
\end{equation}
the gauge potential becomes
\begin{equation}
A
=
-\frac{q}{4\pi}\,\frac{\cos\rho}{\sqrt{H}}\,dT.
\end{equation}
Hence
\begin{equation}
A_\tau=-\frac{q}{4\pi}\cot R\,\partial_\tau T,
\qquad
A_\rho=-\frac{q}{4\pi}\cot R\,\partial_\rho T,
\end{equation}
and therefore
\begin{equation}
F_{\rho\tau}
=
\partial_\rho A_\tau-\partial_\tau A_\rho
=
-\frac{q}{4\pi}
\left[
\partial_\rho(\cot R)\,\partial_\tau T
-
\partial_\tau(\cot R)\,\partial_\rho T
\right].
\end{equation}


We first compute the derivatives of \(T\). From
\begin{equation}
\tan T=\frac{\gamma\Xi}{\cos\tau},
\end{equation}
we have
\begin{equation}
\partial_\tau T
=
\frac{1}{1+\tan^2T}\,
\partial_\tau\!\left(\frac{\gamma\Xi}{\cos\tau}\right).
\end{equation}
Now
\begin{equation}
1+\tan^2T
=
1+\frac{\gamma^2\Xi^2}{\cos^2\tau}
=
\frac{\Delta}{\cos^2\tau},
\end{equation}
while
\begin{equation}
\partial_\tau\Xi=\cos\tau.
\end{equation}
Therefore
\begin{align}
\partial_\tau\!\left(\frac{\gamma\Xi}{\cos\tau}\right)
&=
\gamma\,
\frac{(\partial_\tau\Xi)\cos\tau+\Xi\sin\tau}{\cos^2\tau}
\nonumber\\
&=
\gamma\,\frac{\cos^2\tau+\Xi\sin\tau}{\cos^2\tau}
\nonumber\\
&=
\gamma\,\frac{1-\beta\sin\rho\,\mu\,\sin\tau}{\cos^2\tau},
\end{align}
and hence
\begin{equation}
\partial_\tau T
=
\frac{\gamma\bigl(1-\beta\sin\rho\,\mu\,\sin\tau\bigr)}{\Delta}.
\end{equation}

Similarly, since
\begin{equation}
\partial_\rho\Xi=-\beta\mu\cos\rho,
\end{equation}
we find
\begin{equation}
\partial_\rho\!\left(\frac{\gamma\Xi}{\cos\tau}\right)
=
-\frac{\beta\gamma\mu\cos\rho}{\cos\tau},
\end{equation}
and therefore
\begin{equation}
\partial_\rho T
=
-\frac{\beta\gamma\mu\cos\rho\cos\tau}{\Delta}.
\end{equation}

We next compute the derivatives of \(\cot R\). Since
\begin{equation}
\cot R=\frac{\cos\rho}{\sqrt H},
\end{equation}
we obtain
\begin{equation}
\partial_\tau H
=
-2\sin\tau\cos\tau+2\gamma^2\Xi\cos\tau
=
2\cos\tau(\gamma^2\Xi-\sin\tau),
\end{equation}
so that
\begin{equation}
\partial_\tau(\cot R)
=
-\frac{\cos\rho\cos\tau(\gamma^2\Xi-\sin\tau)}{H^{3/2}}.
\end{equation}
Likewise,
\begin{equation}
\partial_\rho H
=
2\sin\rho\cos\rho-2\beta\gamma^2\mu\cos\rho\,\Xi,
\end{equation}
which gives
\begin{align}
\partial_\rho(\cot R)
&=
-\frac{\sin\rho}{\sqrt H}
-\frac{\cos\rho}{2}H^{-3/2}\partial_\rho H
\nonumber\\
&=
-\frac{\sin\rho\,\Delta-\beta\gamma^2\mu\Xi\cos^2\rho}{H^{3/2}}.
\end{align}
Thus
\begin{equation}
\partial_\rho(\cot R)
=
-\frac{\sin\rho\,\Delta-\beta\gamma^2\mu\Xi\cos^2\rho}{H^{3/2}}.
\end{equation}

\subsection{Global-coordinate expression for \texorpdfstring{$F_{\rho\tau}$}{Frhotau}}
\label{subsec:ads-global-frhotau}

Substituting these derivatives into the general formula for \(F_{\rho\tau}\), we find
\begin{align}
F_{\rho\tau}
&=
-\frac{q}{4\pi}
\left[
\partial_\rho(\cot R)\,\partial_\tau T
-
\partial_\tau(\cot R)\,\partial_\rho T
\right]
\nonumber\\[4pt]
&=
\frac{q\gamma}{4\pi\,\Delta\,H^{3/2}}
\Bigl[
\bigl(\sin\rho\,\Delta-\beta\gamma^2\mu\Xi\cos^2\rho\bigr)
\bigl(1-\beta\sin\rho\,\mu\,\sin\tau\bigr)
\nonumber\\
&\hspace{4cm}
-
\beta\mu\cos^2\rho\cos^2\tau(\gamma^2\Xi-\sin\tau)
\Bigr].
\end{align}
A straightforward simplification using
\begin{equation}
\Xi=\sin\tau-\beta\sin\rho\,\mu,
\qquad
\gamma^2-1=\gamma^2\beta^2,
\end{equation}
shows that the square bracket collapses to
\begin{equation}
\Delta\bigl(\sin\rho-\beta\mu\sin\tau\bigr).
\end{equation}
The factor of \(\Delta\) therefore cancels, and the result takes the compact form
\begin{equation}
F_{\rho\tau}(\tau,\rho,\hat x)
=
\frac{q\gamma\bigl(\sin\rho-\beta\mu\sin\tau\bigr)}
{4\pi\Bigl(\cos^2\tau-\cos^2\rho+\gamma^2\bigl(\sin\tau-\beta\sin\rho\,\mu\bigr)^2\Bigr)^{3/2}}.
\end{equation}
Equivalently, writing \(\beta\mu=\vec\beta\cdot\hat x\),
\begin{equation}
F_{\rho\tau}(\tau,\rho,\hat x)
=
\frac{q\gamma\bigl(\sin\rho-\sin\tau\,\vec\beta\cdot\hat x\bigr)}
{4\pi\Bigl(\cos^2\tau-\cos^2\rho+\gamma^2\bigl(\sin\tau-\sin\rho\,\vec\beta\cdot\hat x\bigr)^2\Bigr)^{3/2}}.
\end{equation}

This is the desired expression for the field strength component in the original global AdS
coordinates. It is the direct analogue of the flat-space Li\'enard--Wiechert Coulombic field,
but now expressed on the global AdS cylinder.


We now evaluate the field near the future AdS null fringe,
\begin{equation}
\rho\to\frac{\pi}{2},
\qquad
\tau\to+\frac{\pi}{2}.
\end{equation}
In this limit,
\begin{equation}
\sin\rho\to 1,
\qquad
\cos\rho\to 0,
\qquad
\sin\tau\to 1,
\qquad
\cos\tau\to 0.
\end{equation}
Therefore the numerator approaches
\begin{equation}
q\gamma\bigl(1-\vec\beta\cdot\hat x\bigr),
\end{equation}
while the denominator becomes
\begin{equation}
4\pi
\Bigl[
\gamma^2\bigl(1-\vec\beta\cdot\hat x\bigr)^2
\Bigr]^{3/2}
=
4\pi\,\gamma^3\bigl(1-\vec\beta\cdot\hat x\bigr)^3.
\end{equation}
Hence
\begin{equation}
\lim_{\substack{\rho\to\pi/2\\ \tau\to+\pi/2}}
F_{\rho\tau}(\tau,\rho,\hat x)
=
\frac{q}{4\pi\,\gamma^2\bigl(1-\vec\beta\cdot\hat x\bigr)^2}.
\end{equation}
This is the leading Coulombic field strength at the future null fringe.


We next approach the past AdS null fringe,
\begin{equation}
\rho\to\frac{\pi}{2},
\qquad
\tau\to-\frac{\pi}{2}.
\end{equation}
Now
\begin{equation}
\sin\rho\to 1,
\qquad
\cos\rho\to 0,
\qquad
\sin\tau\to -1,
\qquad
\cos\tau\to 0,
\end{equation}
so that the numerator tends to
\begin{equation}
q\gamma\bigl(1+\vec\beta\cdot\hat x\bigr),
\end{equation}
and the denominator becomes
\begin{equation}
4\pi
\Bigl[
\gamma^2\bigl(1+\vec\beta\cdot\hat x\bigr)^2
\Bigr]^{3/2}
=
4\pi\,\gamma^3\bigl(1+\vec\beta\cdot\hat x\bigr)^3.
\end{equation}
Therefore
\begin{equation}
\lim_{\substack{\rho\to\pi/2\\ \tau\to-\pi/2}}
F_{\rho\tau}(\tau,\rho,\hat x)
=
\frac{q}{4\pi\,\gamma^2\bigl(1+\vec\beta\cdot\hat x\bigr)^2}.
\end{equation}
This is the Coulombic field strength at the past null fringe.

\subsection{Antipodal matching}
\label{subsec:ads-antipodal-matching}

We can now compare the two fringe limits. At fixed angle \(\hat x\), the future and past
limits are
\begin{equation}
\lim_{\substack{\rho\to\pi/2\\ \tau\to+\pi/2}}
F_{\rho\tau}(\tau,\rho,\hat x)
=
\frac{q}{4\pi\,\gamma^2\bigl(1-\vec\beta\cdot\hat x\bigr)^2},
\end{equation}
and
\begin{equation}
\lim_{\substack{\rho\to\pi/2\\ \tau\to-\pi/2}}
F_{\rho\tau}(\tau,\rho,\hat x)
=
\frac{q}{4\pi\,\gamma^2\bigl(1+\vec\beta\cdot\hat x\bigr)^2}.
\end{equation}
These are not equal when compared at the same angle. However, under the antipodal map
\begin{equation}
\hat x\longmapsto -\hat x,
\end{equation}
one has
\begin{equation}
\vec\beta\cdot(-\hat x)=-(\vec\beta\cdot\hat x),
\end{equation}
and therefore
\begin{equation}
\lim_{\substack{\rho\to\pi/2\\ \tau\to-\pi/2}}
F_{\rho\tau}(\tau,\rho,-\hat x)
=
\frac{q}{4\pi\,\gamma^2\bigl(1-\vec\beta\cdot\hat x\bigr)^2}.
\end{equation}
This is exactly the future null-fringe value. We thus obtain the antipodal matching relation
\begin{equation}
\lim_{\substack{\rho\to\pi/2\\ \tau\to+\pi/2}}
F_{\rho\tau}(\tau,\rho,\hat x)
=
\lim_{\substack{\rho\to\pi/2\\ \tau\to-\pi/2}}
F_{\rho\tau}(\tau,\rho,-\hat x).
\end{equation}

This is the AdS null-fringe version of the usual flat-space antipodal matching law. Its
geometric meaning is clear: the natural continuity statement is not between equal angles on
the two boundary spheres at \(\tau=\pm \pi/2\), but rather between antipodal points, reflecting
the projective null structure of the AdS boundary. In the large-\(L\) limit, this reduces
precisely to the familiar matching condition for the leading Coulombic data at
\(\mathscr I^+_-\) and \(\mathscr I^-_+\).

\paragraph{Antipodal matching of \(F_{\rho\tau}\).}
In global coordinates the AdS antipodal map acts as
\begin{equation}
\mathscr A:\qquad
\tau\to \tau\pm\pi,\qquad
\rho\to \rho,\qquad
\Omega\to \Omega_A,
\end{equation}
where \(\Omega_A\) is the antipodal point on \(S^2\). Since
\begin{equation}
\hat x(\Omega_A)=-\hat x(\Omega),
\end{equation}
it follows that
\begin{equation}
\vec\beta\cdot \hat x(\Omega_A)= -\,\vec\beta\cdot \hat x(\Omega).
\end{equation}
Now consider
\begin{equation}
F_{\rho\tau}(\tau,\rho,\Omega)
=
\frac{q\gamma\bigl(\sin\rho-\sin\tau\,\vec\beta\cdot \hat x(\Omega)\bigr)}
{4\pi\Bigl(\cos^2\tau-\cos^2\rho+\gamma^2\bigl(\sin\tau-\sin\rho\,\vec\beta\cdot \hat x(\Omega)\bigr)^2\Bigr)^{3/2}}.
\end{equation}
Under \(\mathscr A\),
\begin{equation}
\sin(\tau\pm\pi)=-\sin\tau,
\qquad
\cos^2(\tau\pm\pi)=\cos^2\tau,
\end{equation}
and therefore the numerator transforms as
\begin{align}
\sin\rho-\sin(\tau\pm\pi)\,\vec\beta\cdot \hat x(\Omega_A)
&=
\sin\rho-(-\sin\tau)\,(-\vec\beta\cdot \hat x(\Omega))
\nonumber\\
&=
\sin\rho-\sin\tau\,\vec\beta\cdot \hat x(\Omega),
\end{align}
while the nontrivial denominator factor becomes
\begin{align}
\sin(\tau\pm\pi)-\sin\rho\,\vec\beta\cdot \hat x(\Omega_A)
&=
-\sin\tau+\sin\rho\,\vec\beta\cdot \hat x(\Omega)
\nonumber\\
&=
-\bigl(\sin\tau-\sin\rho\,\vec\beta\cdot \hat x(\Omega)\bigr).
\end{align}
After squaring, this is unchanged. Hence the full denominator is invariant, and we conclude
\begin{equation}
F_{\rho\tau}(\tau\pm\pi,\rho,\Omega_A)=F_{\rho\tau}(\tau,\rho,\Omega).
\end{equation}
Since \(\partial \tau'/\partial \tau=1\) and \(\partial \rho'/\partial \rho=1\), there is no Jacobian sign in this component, so equivalently
\begin{equation}
(\mathscr A^*F)_{\rho\tau}=F_{\rho\tau}.
\end{equation}

The global AdS bulk solution satisfies a stronger statement than the usual flat-space matching law, namely the exact antipodal covariance
\[
F_{\rho\tau}(\tau+\pi,\rho,\Omega_A)=F_{\rho\tau}(\tau,\rho,\Omega),
\]
valid for arbitrary bulk radius \(\rho\). This is an intrinsic consequence of the AdS antipodal isometry. However, the flat-space infrared matching problem concerns not the full bulk field but the leading Coulombic asymptotic data on null infinity. To isolate those data in global AdS, one must first approach the conformal boundary \(\rho\to \pi/2\), and then focus on the distinguished boundary times \(\tau\to \pm \pi/2\), which are selected by radial null propagation from the bulk center. These two null-fringe regions are the finite-\(L\) precursors of \(\mathscr I^\pm\), and in the large-\(L\) limit their antipodal identification reduces precisely to the standard flat-space matching condition at \(\mathscr I^+_-\) and \(\mathscr I^-_+\).

\section{Embedding-space origin of the AdS antipodal map}
\label{sec:ads-antipodal-map}
A geometric way to understand the AdS antipodal map is to view Lorentzian
\(\mathrm{AdS}_4\) as the hyperboloid
\begin{equation}
X\cdot X=-L^2
\end{equation}
embedded in \(\mathbb R^{3,2}\) with metric
\begin{equation}
\eta_{AB}=\mathrm{diag}(-1,-1,+1,+1,+1).
\end{equation}
The antipodal map is simply the central inversion
\begin{equation}
\mathscr A:\qquad X^A\mapsto -X^A.
\end{equation}
Since
\begin{equation}
(-X)\cdot(-X)=X\cdot X=-L^2,
\end{equation}
this map preserves the AdS quadric, and because it is induced by a linear
\(O(3,2)\) transformation, it is an isometry of AdS.

To read this map in global coordinates, recall the standard parametrization
\begin{equation}
X_1=\frac{L\cos\tau}{\cos\rho},
\qquad
X_2=\frac{L\sin\tau}{\cos\rho},
\qquad
(X_3,X_4,X_5)=L\tan\rho\,\hat x(\Omega).
\end{equation}
Under \(X\mapsto -X\), the spatial part becomes
\begin{equation}
(X_3,X_4,X_5)\mapsto -(X_3,X_4,X_5),
\end{equation}
so the unit vector on the sphere transforms as
\begin{equation}
\hat x(\Omega)\mapsto -\hat x(\Omega)=\hat x(\Omega_A),
\end{equation}
namely
\begin{equation}
\Omega\mapsto \Omega_A.
\end{equation}
Moreover, \(\rho\) is unchanged, because it depends only on the squares of the
embedding coordinates:
\begin{equation}
X_3^2+X_4^2+X_5^2=L^2\tan^2\rho,
\qquad
X_1^2+X_2^2=\frac{L^2}{\cos^2\rho}.
\end{equation}
Finally, the temporal components transform as
\begin{equation}
X_1\mapsto -X_1,
\qquad
X_2\mapsto -X_2,
\end{equation}
which implies
\begin{equation}
\cos\tau\mapsto -\cos\tau,
\qquad
\sin\tau\mapsto -\sin\tau.
\end{equation}
Hence
\begin{equation}
\tau\mapsto \tau+\pi \qquad (\mathrm{mod}\;2\pi).
\end{equation}
On the universal cover this is written as
\begin{equation}
\tau\mapsto \tau\pm\pi,
\end{equation}
with the sign corresponding to the choice of lift. Therefore in global coordinates
the antipodal map acts as
\begin{equation}
\mathscr A:\qquad
\tau\to \tau\pm\pi,\qquad
\rho\to \rho,\qquad
\Omega\to \Omega_A.
\end{equation}

This also has a direct geodesic interpretation. A timelike geodesic in embedding
space is
\begin{equation}
Y(s)=P\cos\frac{s}{L}+LU\sin\frac{s}{L},
\end{equation}
and one immediately finds
\begin{equation}
Y(s+\pi L)=-Y(s).
\end{equation}
Thus the antipodal map sends each point on a timelike geodesic to the point
separated from it by proper time \(\pi L\). On the central geodesic, where
\(s=L\tau\), this reduces precisely to \(\tau\to\tau+\pi\).

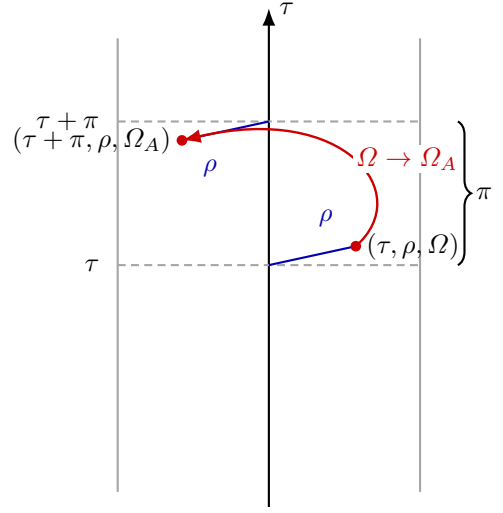
\begin{figure}[H]
\centering
\begin{tikzpicture}[>=Latex,thick,font=\small]

\begin{scope}[xshift=0cm]

\node at (0,4.1) {(a) Embedding-space origin of the antipodal map};

\draw[->] (-3.2,0) -- (3.2,0) node[below] {$X_{\mathrm{space}}$};
\draw[->] (0,-3.2) -- (0,3.4) node[left] {$X_{\mathrm{time}}$};

\draw[blue!70!black,very thick,domain=-2.2:2.2,samples=120,smooth]
plot (\x,{sqrt(1.4*1.4 + \x*\x)});
\draw[blue!70!black,very thick,domain=-2.2:2.2,samples=120,smooth]
plot (\x,{-sqrt(1.4*1.4 + \x*\x)});

\node[blue!70!black] at (2.05,3.05) {$X\cdot X=-L^2$};

\fill (0,0) circle (1.3pt);
\node[below right] at (0,0) {$0$};

\coordinate (Xp) at (1.45,2.02);
\coordinate (Xm) at (-1.45,-2.02);
\draw[dashed,gray!70] ($(Xm)+(-0.45,-0.63)$) -- ($(Xp)+(0.45,0.63)$);

\fill[red!80!black] (Xp) circle (2pt);
\fill[red!80!black] (Xm) circle (2pt);
\node[right] at (Xp) {$X$};
\node[left] at (Xm) {$-X$};

\draw[->,red!80!black,line width=0.9pt]
(Xp) -- node[below left=1pt,fill=white,inner sep=1pt] {$\mathscr A:X\mapsto -X$} (Xm);

\node[align=center,font=\scriptsize] at (0,-3.95)
{same ambient line through the origin\\
intersects the AdS hyperboloid at antipodal points};

\end{scope}

\begin{scope}[xshift=9cm]

\node at (0,4.1) {(b) Global AdS cylinder};

\draw[gray!70] (-2.0,-3.0) -- (-2.0,3.0);
\draw[gray!70] ( 2.0,-3.0) -- ( 2.0,3.0);

\draw[->] (0,-3.2) -- (0,3.4) node[right] {$\tau$};

\draw[gray!70,densely dashed] (-2.0,0.0) -- (2.0,0.0);
\draw[gray!70,densely dashed] (-2.0,1.9) -- (2.0,1.9);
\node[left] at (-2.1,0.0) {$\tau$};
\node[left] at (-2.1,1.9) {$\tau+\pi$};

\node[font=\scriptsize] at (0,-3.95) {$\rho=\frac{\pi}{2}$: conformal boundary};

\coordinate (C1) at (0,0.0);
\coordinate (C2) at (0,1.9);
\coordinate (P)  at (1.15,0.25);
\coordinate (PA) at (-1.15,1.65);

\draw[blue!70!black] (C1) -- (P);
\draw[blue!70!black] (C2) -- (PA);
\node[blue!70!black] at (0.75,0.62) {$\rho$};
\node[blue!70!black] at (-0.78,1.28) {$\rho$};

\fill[red!80!black] (P) circle (2pt);
\fill[red!80!black] (PA) circle (2pt);
\node[right] at (P) {$({\tau},\rho,\Omega)$};
\node[left]  at (PA) {$({\tau}+\pi,\rho,\Omega_A)$};

\draw[->,red!80!black,line width=0.9pt]
(P) .. controls (1.95,0.95) and (1.0,2.15) ..
node[right,fill=white,inner sep=1pt] {$\Omega\to\Omega_A$}
(PA);

\draw[decorate,decoration={brace,mirror,amplitude=5pt}] (2.50,0.0) -- (2.50,1.9);
\node[right] at (2.6,0.95) {$\pi$};

\end{scope}

\end{tikzpicture}
\caption{Geometric visualization of the AdS antipodal map. Left: in embedding space, the antipodal map is the central inversion \(X\mapsto -X\) on the hyperboloid \(X\cdot X=-L^2\). Right: in global coordinates it becomes \((\tau,\rho,\Omega)\mapsto(\tau+\pi,\rho,\Omega_A)\), namely a shift by \(\pi\) in global time together with the antipodal map on the sphere, at fixed radial coordinate \(\rho\).}
\label{fig:ads-antipodal-map}
\end{figure}

\section{Image charges, boundary conditions, and antipodal data in AdS}
\label{sec:ads_image_charge}

The AdS Coulomb seed also admits a useful image-charge interpretation.  This
interpretation is closely related to, but conceptually distinct from, the flat-space image
charge at spatial infinity.  In flat-space, after conformal compactification, a spatial slice
becomes a compact \(S^3\) without boundary.  Gauss's law then forbids a smooth Maxwell
configuration with nonzero total charge, so the Coulomb field of an isolated charge must be
represented on the compactified slice by a compensating image singularity at \(i^0\).  In
AdS the logic is different.  A constant global-time slice of \(\mathrm{AdS}_4\), after stripping
off the conformal factor, is not a closed \(S^3\), but rather a hemisphere of \(S^3\) with
boundary.  The image charge therefore does not arise from a zero-mode obstruction on a
closed compact space.  Instead, it encodes the boundary condition imposed at the timelike
AdS conformal boundary.

To make this precise, recall that the global \(\mathrm{AdS}_4\) metric is
\begin{equation}
\label{eq:ads_metric_image_section}
ds^2
=
\frac{L^2}{\cos^2\rho}
\left(
-d\tau^2+d\rho^2+\sin^2\rho\,d\Omega_2^2
\right),
\qquad
0\leq \rho<\frac{\pi}{2}.
\end{equation}
Since four-dimensional Maxwell theory is conformally invariant, the electrostatic problem
may equivalently be studied in the conformally related geometry
\begin{equation}
\label{eq:half_esu_image_section}
d\widetilde s^2
=
-d\tau^2+d\rho^2+\sin^2\rho\,d\Omega_2^2,
\qquad
0\leq \rho\leq\frac{\pi}{2}.
\end{equation}
Thus a constant-\(\tau\) slice is the spatial hemisphere
\begin{equation}
\label{eq:ads_hemisphere_image_section}
\Sigma_+
=
\left\{
0\leq \rho\leq \frac{\pi}{2}
\right\}
\subset S^3,
\qquad
\partial\Sigma_+ \simeq S^2 .
\end{equation}
The equator \(\rho=\pi/2\) is the conformal boundary of AdS.

The static Coulomb seed obtained above is
\begin{equation}
\label{eq:ads_static_seed_image_section}
A
=
-\frac{q}{4\pi}\cot\rho\,d\tau .
\end{equation}
It is useful to write this as
\begin{equation}
\label{eq:ads_scalar_potential_image_section}
A=-\Phi(\rho)\,d\tau,
\qquad
\Phi(\rho)=\frac{q}{4\pi}\cot\rho .
\end{equation}
The additive constant in the electrostatic potential has been fixed so that
\begin{equation}
\label{eq:ads_dirichlet_image_section}
\Phi\big|_{\rho=\pi/2}=0 .
\end{equation}
Therefore the AdS Coulomb seed is naturally the Dirichlet Green function on the hemisphere
\(\Sigma_+\).  Indeed, away from the source at \(\rho=0\), the scalar Laplacian on the unit
hemisphere,
\begin{equation}
d\ell^2_{\Sigma_+}
=
d\rho^2+\sin^2\rho\,d\Omega_2^2 ,
\end{equation}
acts as
\begin{equation}
\label{eq:laplacian_cot_image_section}
\Delta_{\Sigma_+}\cot\rho
=
\frac{1}{\sin^2\rho}
\partial_\rho
\left(
\sin^2\rho\,\partial_\rho\cot\rho
\right)
=
0,
\qquad
\rho\neq0 .
\end{equation}
Distributionally, however, \(\Phi\) has a Coulomb singularity at the north pole
\(N\equiv \rho=0\).  With Dirichlet boundary condition at \(\partial\Sigma_+\), the precise
Green-function equation is
\begin{equation}
\label{eq:dirichlet_green_hemisphere_image_section}
-\Delta_{\Sigma_+}\Phi
=
q\,\delta^{(3)}_{\Sigma_+}(x,N),
\qquad
\Phi\big|_{\partial\Sigma_+}=0 .
\end{equation}
Equivalently, for any smooth test function \(f\) vanishing on the boundary,
\begin{equation}
\label{eq:weak_dirichlet_image_section}
\int_{\Sigma_+}
\nabla_a\Phi\,\nabla^a f\,
\mathrm{vol}_{\Sigma_+}
=
q\,f(N).
\end{equation}
This is the precise sense in which the static AdS Coulomb field is a Dirichlet Green
function on the spatial hemisphere.

The image-charge interpretation becomes manifest after doubling the hemisphere
\(\Sigma_+\) across its boundary.  Let
\begin{equation}
\label{eq:doubled_hemisphere_image_section}
D\Sigma_+\simeq S^3
\end{equation}
denote the doubled spatial slice, obtained by reflecting
\begin{equation}
\label{eq:rho_reflection_image_section}
\rho\longmapsto \pi-\rho .
\end{equation}
The AdS boundary \(\rho=\pi/2\) becomes the equatorial fixed surface of this reflection.
The Dirichlet Green function on the hemisphere is then obtained by restricting to
\(0\leq \rho\leq \pi/2\) the odd Green function on the full \(S^3\),
\begin{equation}
\label{eq:odd_green_s3_image_section}
\Phi_D(\rho)
=
\frac{q}{4\pi}\cot\rho,
\qquad
0<\rho<\pi .
\end{equation}
This doubled potential satisfies
\begin{equation}
\label{eq:s3_image_charge_equation_section}
-\Delta_{S^3}\Phi_D
=
q\,\delta^{(3)}_{S^3}(x,N)
-
q\,\delta^{(3)}_{S^3}(x,S),
\end{equation}
where \(S\) is the point antipodal to \(N\) on the doubled \(S^3\), namely \(\rho=\pi\).
Thus the Dirichlet AdS Coulomb field can be viewed as the field of a physical charge
\(+q\) at the AdS center together with an image charge \(-q\) in the reflected copy beyond
the AdS boundary.

This is the direct AdS analogue of the elementary electrostatic image-charge method.
The AdS boundary plays the role of a reflecting surface for the Dirichlet representative of
the scalar potential.  The condition \(\Phi|_{\rho=\pi/2}=0\) is enforced by placing an
opposite image charge in the second copy of the doubled geometry.  The local behavior of
the doubled potential makes this interpretation explicit
\begin{equation}
\label{eq:source_singularity_image_section}
\Phi_D(\rho)
\sim
\frac{q}{4\pi\rho},
\qquad
\rho\to0,
\end{equation}
near the physical charge, while
\begin{equation}
\label{eq:image_singularity_image_section}
\Phi_D(\rho)
\sim
-\frac{q}{4\pi(\pi-\rho)},
\qquad
\rho\to\pi,
\end{equation}
near the image charge.  The equator lies halfway between the two singularities, and there
\begin{equation}
\label{eq:equator_vanishes_image_section}
\Phi_D\left(\frac{\pi}{2}\right)=0 .
\end{equation}
The image charge is therefore not an additional physical source in the AdS bulk.  It is a
bookkeeping device which replaces the boundary condition by an auxiliary source in the
doubled geometry.

There is also a useful flux interpretation.  The field strength of the static seed is
\begin{equation}
\label{eq:ads_seed_field_strength_image_section}
F=dA
=
\frac{q}{4\pi}\csc^2\rho\,d\rho\wedge d\tau .
\end{equation}
On the physical hemisphere, the normal electric field at the AdS boundary is
\begin{equation}
\label{eq:boundary_flux_density_image_section}
\mathscr E_{\partial}
=
-\left.\partial_\rho\Phi\right|_{\rho=\pi/2}
=
\frac{q}{4\pi}.
\end{equation}
Hence the total flux through the boundary sphere is
\begin{equation}
\label{eq:boundary_flux_total_image_section}
\int_{\partial\Sigma_+}
\mathscr E_{\partial}\,d\Omega_2
=
q .
\end{equation}
Thus, in the physical AdS spacetime, the electric flux sourced by the bulk charge reaches
the timelike conformal boundary.  In the doubled description, the same field lines continue
through the equator into the reflected copy and terminate on the image charge \(-q\).  The
two descriptions are equivalent: the physical AdS description uses a boundary condition,
whereas the doubled description replaces that boundary condition by an image source.

This image-charge picture extends directly to the geodesic-centered construction of the
AdS Li\'enard--Wiechert field.  Let \((T,R,\Theta,\Phi)\) be coordinates adapted to a
timelike geodesic \(\Gamma\).  In this frame the moving charge is at rest at \(R=0\), and the
field is simply the AdS Coulomb seed
\begin{equation}
\label{eq:geodesic_seed_image_section}
A
=
-\frac{q}{4\pi}\cot R\,dT,
\qquad
F
=
\frac{q}{4\pi}\csc^2R\,dR\wedge dT .
\end{equation}
The corresponding spatial slice is again a hemisphere, now centered on the source geodesic.
After doubling this adapted hemisphere across its boundary, the same solution may be
viewed as the field of a charge \(+q\) on the physical geodesic \(\Gamma\), together with an
image charge \(-q\) on the reflected worldline behind the AdS boundary.  Rewriting the
adapted variables \((T,R)\) in terms of the original global coordinates
\((\tau,\rho,\Omega)\) gives the AdS Li\'enard--Wiechert field.  From this viewpoint, the
moving solution is not a new dynamical configuration obtained by solving Maxwell's
equations again; it is the Coulomb field of the charge, together with its boundary image,
written in a frame not centered on the source geodesic.

The image-charge interpretation should be distinguished from the antipodal covariance of
the AdS field.  The image charge is tied to the choice of boundary condition at the timelike
AdS boundary.  It explains how the Coulomb seed can carry nonzero electric flux while the
Dirichlet representative of the potential vanishes at \(\rho=\pi/2\).  The antipodal
covariance, by contrast, is a statement about the global AdS isometry
\begin{equation}
\label{eq:antipodal_map_image_section}
X^A\mapsto -X^A,
\qquad
(\tau,\rho,\Omega)\mapsto(\tau+\pi,\rho,\Omega_A).
\end{equation}
This map explains why the Coulombic data extracted at the future and past AdS null
fringes are related by antipodal identification.  Thus the image charge accounts for the
boundary realization of the Coulomb seed, while the antipodal map accounts for the
matching of its asymptotic null-fringe data.

In this sense, the image charge is the global bookkeeping device which tracks how a local
Coulomb field is completed once the spatial geometry is either compactified or supplied with
a boundary.  In flat-space, conformal compactification turns the escaping Coulomb flux into
an image singularity at spatial infinity \(i^0\).  In AdS, the timelike conformal boundary
is represented by the equator of a spatial hemisphere, and the Dirichlet Coulomb seed is
equivalently described by an opposite image charge in the reflected copy of the doubled
geometry.  In the large-\(L\) limit, these two viewpoints merge: the AdS boundary recedes
to infinity, the boundary image degenerates into the familiar Coulombic singular structure
at spatial infinity, and the AdS null-fringe antipodal relation becomes the standard
flat-space matching law between \(\mathscr I^+_-\) and \(\mathscr I^-_+\).

\section{Conclusions and outlook}
\label{sec:conclusions}

In this paper we have presented a geometric derivation of Li\'enard--Wiechert fields in $\mathrm{AdS}_4$ and clarified how the antipodal matching condition of flat-space electrodynamics emerges from the AdS perspective. The central idea was to identify the correct AdS analogue of uniform inertial motion. In Minkowski space, a uniformly moving charge is naturally described relative to the timelike geodesic that defines its worldline, and in the corresponding rest frame its field is simply the ordinary Coulomb solution. Our analysis showed that the familiar Li\'enard--Wiechert field is therefore most naturally understood not as a fundamentally new solution, but as the Coulomb field rewritten in coordinates that are not centered on the source trajectory. This viewpoint makes the angular dependence of the asymptotic field and the associated matching law at null infinity conceptually transparent.

We then extended this logic to global $\mathrm{AdS}_4$, where the natural replacement of inertial motion is motion along a timelike geodesic. Instead of constructing the field of a moving charge by acting with explicit $\mathrm{SO}(3,2)$ isometries on the static solution, we began from the static Coulomb problem for a charge sitting at the center of AdS. We then introduced coordinates adapted to an arbitrary timelike geodesic, in which the same freely moving charge is again at rest, and finally reconstructed the solution in arbitrary global coordinates using embedding-space invariants. In this way, the AdS Li\'enard--Wiechert field acquires a simple geometric interpretation: it is the central Coulomb seed viewed from a frame recentered on the relevant geodesic. The moving solution is therefore not obtained by solving Maxwell's equations anew, but by expressing one and the same Coulombic configuration in different geodesically adapted descriptions.

A key outcome of this construction is that the AdS bulk field exhibits an exact antipodal covariance already at finite radius. This statement is a direct consequence of the embedding-space antipodal map and is therefore intrinsic to the global geometry of AdS, rather than an artifact of taking an asymptotic limit. The usual flat-space antipodal matching law does not arise from this bulk relation alone; rather, it appears after restricting to the conformal boundary and, more specifically, to the distinguished boundary times selected by radial null propagation from the AdS center. These two null-fringe regions are the finite-$L$ precursors of $\mathscr I^\pm$, and the leading Coulombic field evaluated there obeys precisely the matching relation that becomes, in the large-$L$ limit, the standard antipodal identification between the data at $\mathscr I^+_-$ and $\mathscr I^-_+$. In this sense, the flat-space infrared matching condition is naturally reinterpreted as the large-radius remnant of a more primitive geometric structure already present in AdS.

Our derivation therefore provides a unified picture of Li\'enard--Wiechert fields in flat-space and in AdS. It emphasizes that the physically relevant object is a geodesic observable: the electromagnetic field sourced by a charge in uniform motion is the static Coulomb field expressed in coordinates centered on the appropriate source geodesic. From this perspective, the asymptotic matching relation is not an additional dynamical input but a geometric consequence of how null propagation, conformal infinity, and antipodal structure are organized in the global AdS description. This also sharpens the interpretation of the flat-space limit itself: the matching law at null infinity is inherited from the projective null structure of the AdS boundary together with the distinguished role of the AdS null fringes.

There is also a complementary image-charge interpretation of the static AdS Coulomb seed. After conformal rescaling, a constant global-time slice of AdS is a hemisphere of \(S^3\), whose equator is the timelike AdS boundary. The potential
$A=-\frac{q}{4\pi}\cot\rho\,d\tau$
is the Dirichlet Green function on this hemisphere. Upon doubling the hemisphere across the AdS boundary, this field can be viewed as the field of a physical charge at the AdS center together with an opposite image charge in the reflected copy. The image charge is not an additional physical source; it encodes the boundary condition at \(\rho=\pi/2\). In geodesic-adapted coordinates, the same perspective applies to the moving solution: the AdS Li\'enard--Wiechert field is the Coulomb field of the charge together with its boundary image, rewritten in coordinates centered on the source geodesic. Thus the image-charge picture explains the boundary-condition aspect of the Coulomb seed, while the antipodal covariance explains how its null-fringe data are related in the flat-space limit.


Recent developments suggest that antipodal matching should be viewed not as an auxiliary boundary condition, but as part of the geometric mechanism by which flat-space observables emerge from AdS. In the celestial description, Lorentzian AdS Witten diagrams reduce in the large-radius limit to celestial amplitudes when the boundary operators are inserted on infinitesimal strips around the global-time slices $\tau=\pm \frac{\pi}{2}$, with these two boundary spheres identified antipodally \cite{Duary:2022onm, Duary:2024fii, Duary:2024cqb}; in this limit, the AdS bulk-to-boundary propagators become flat-space conformal primary wavefunctions, so the Witten diagrams reorganize directly into flat-space amplitudes written in a conformal primary basis \cite{deGioia:2024yne,deGioia:2022fcn,Bagchi:2023fbj,Alday:2024yyj,Navarro:2025xln}. For the connection to AdS/CFT, see the review \cite[Sec.~9]{Ruzziconi:2026bix} and \cite[Sec.~9]{Zhu:2026ofh}. From the complementary Carrollian perspective, taking the flat-space limit in AdS Bondi coordinates makes the bulk-to-flat reduction manifest already in position space: the AdS metric smoothly approaches the Minkowski metric, the boundary conformal geometry contracts to a Carrollian geometry at null infinity, and the flat-space limit of Witten diagrams yields Carrollian amplitudes with the natural identification of past and future null infinity inherited from the null geometry \cite{Alday:2024yyj, Li:2024kbo}. More recently, it has also been emphasized that in Lorentzian AdS the relevant null-foliated bulk-to-boundary propagators, their shadow transforms, and the associated boost eigenfunctions furnish a framework in which the flat-space limit leads directly to the Carrollian expansion of bulk fields, while AdS boost eigenstates reduce to massless conformal primary wavefunctions \cite{Navarro:2025xln}. Our AdS antipodal relation fits naturally into this broader picture: the exact bulk covariance under $\Omega\to\Omega_{A}$ and $\tau\to\tau\pm\pi$ may be regarded as the finite-radius precursor of the standard flat-space matching law, while the null-fringe limits near $\tau=\pm \frac{\pi}{2}$ furnish the corresponding in/out data whose equality is restored precisely after antipodal identification. In this sense, the familiar matching condition at $\mathscr I^{+}_{-}$ and $\mathscr I^{-}_{+}$ is naturally interpreted as the large-radius shadow of an antipodal structure already present in AdS.

A further perspective on our result comes from the Lorentzian bulk-point limit of AdS correlators \cite{Maldacena:2015iua,Lam:2017ofc}. The same boundary regions near \(\tau=\pm \frac{\pi}{2}\) that arise here from radial null propagation from the AdS center are precisely the regions singled out in the bulk-point regime, where a local Witten diagram is controlled by a neighborhood of a single bulk interaction point and reorganizes, in the large-radius limit, into the corresponding flat-space amplitude written in the conformal basis. From this viewpoint, the AdS Li\'enard--Wiechert field provides a particularly simple classical probe of the same AdS-to-flat mechanism: the null fringes carry both the in/out data relevant for bulk-point scattering and the leading Coulombic data of the moving charge. The appearance of antipodal matching on precisely these strips is therefore fully consistent with the bulk-point picture and suggests that the infrared matching law should be understood as part of the same geometric reorganization by which flat-space observables emerge from global AdS. At the same time, as in the standard discussion of bulk-point singularities, one should remember that the strict singular behavior is a feature of local bulk perturbation theory and is smoothed by finite \(\alpha'\) effects and, more generally, is not expected to survive as an exact nonperturbative singularity at finite \(G_N\) \cite{Maldacena:2015iua}.

A closely related Lorentzian perspective on the flat-space limit has recently been developed in \cite{Berenstein:2025tts,Berenstein:2025qhb,Adamo:2025bfr}. In particular, \cite{Berenstein:2025tts} emphasizes that the bulk flat-space limit should be understood in terms of Lorentzian state preparation, an In\"on\"u--Wigner contraction of the AdS symmetry algebra, and boosted AdS wavefunctions that become the appropriate flat-space particle states. From this viewpoint, the emergence of flat-space physics is controlled not only by local flatness near a chosen bulk point, but also by how globally defined AdS wave packets reorganize into in/out states in the large-radius limit. The extension of this embedding-space program to higher-spin fields in \cite{Berenstein:2025qhb} further suggests that the geometric strategy adopted here should have natural analogues beyond Maxwell theory: massive spinning primaries admit a clean flat-space limit, while states that become massless involve additional subtleties concentrated in longitudinal polarizations. From a complementary boundary-based perspective, \cite{Adamo:2025bfr} shows in AdS-Bondi coordinates that tree-level AdS correlators in cubic scalar theories pass directly to Carrollian amplitudes in the flat-space limit, making the AdS-to-flat reorganization manifest already in position space. Our AdS Li\'enard--Wiechert construction fits naturally into this broader picture. The null-fringe regions near \(\tau=\pm \pi/2\), singled out here by radial null propagation from the bulk center, play the role of classical carriers of the in/out data that survive the flat-space limit, while the resulting antipodal relation may be viewed as the infrared shadow of the same Lorentzian mechanism by which global AdS observables reorganize into flat-space scattering data.

The analysis of this paper suggests several natural directions for further study. 

\paragraph{Accelerated worldlines, radiation, and memory in AdS.} The most immediate extension is to go beyond geodesic motion and treat genuinely accelerated worldlines in AdS. In flat-space, the full Li\'enard--Wiechert field naturally separates into Coulombic and radiative pieces, and it would be interesting to understand in detail how this decomposition is reorganized by the global AdS geometry and how radiative data are encoded near the AdS null fringes. Closely related is the problem of deriving an AdS version of electromagnetic memory and of understanding whether the asymptotic data at the two fringes admit a formulation directly analogous to the infrared triangle of soft theorems, asymptotic symmetries, and memory in asymptotically flat spacetime. Recent work on electromagnetic multipole expansions near spatial infinity has uncovered infinite families of antipodal matching relations and clarified their connection to both the leading and logarithmic soft photon theorems \cite{Compere:2025tzr}.\footnote{In particular, Ref.~\cite{Compere:2025tzr} derives antipodal matching relations for the electromagnetic field near spatial infinity, including Coulombic tail corrections, and uses the resulting relations to rederive the classical logarithmic soft photon theorem. It would be interesting to determine whether analogous structures arise from the two AdS fringes in the flat-space limit.} A complementary finite-distance formulation of gravitational memory in terms of Carrollian null screens has recently been developed in \cite{Diaz:2026igh}. It would be interesting to understand whether this framework admits an AdS counterpart and how its flat-space limit connects to the finite-distance Carrollian memory construction. Such a relation could provide a geometric bridge between electromagnetic and gravitational memory in AdS and the corresponding infrared structures of asymptotically flat spacetime.

\paragraph{Leaky boundary conditions and \(\Lambda\)-BMS structure.}
A further direction is to understand how the above Coulombic and radiative data depend on the choice of boundary condition at the AdS boundary. Standard reflecting boundary conditions in global AdS do not describe an open scattering problem: radiation reaching the timelike boundary is reflected back into the bulk. For questions closer to flat-space scattering, soft radiation, and memory, it is natural to consider more general, ``leaky'' or flux-permitting boundary conditions, in which part of the radiative data is allowed to pass through the boundary. This is closely related to the \(\Lambda\)-BMS perspective on asymptotically locally \(\mathrm{(A)dS}_4\) spacetimes. In the gravitational case, the residual symmetry group after boundary gauge fixing is the \(\Lambda\)-BMS\(_4\) group, which reduces to the extended BMS\(_4\) group in the asymptotically flat limit \(\Lambda\to0\) \cite{Compere:2019bua}. The relevant boundary flux is controlled by the pair \((q_{AB},J^{AB})\), which plays the role of the \(\mathrm{(A)dS}_4\) analogue of Bondi shear/news data. In AdS, imposing boundary conditions such as \(J^{AB}=0\) removes the symplectic flux and reduces the allowed asymptotic symmetry group, whereas allowing nonzero flux suggests a more open, radiative phase space. It would therefore be interesting to formulate an electromagnetic analogue of these leaky \(\Lambda\)-BMS boundary conditions and to ask whether the AdS null-fringe data of accelerated charges realize, in the large-radius limit, the usual flat-space infrared triangle of soft theorems, asymptotic symmetries, and memory.

\paragraph{Extension to gravitational fields.}
An important direction is to extend the present geodesic-centered construction from Maxwell theory to gravity. Since the central idea of this work is fundamentally geometric-namely, that the field of a uniformly moving source in AdS is most naturally understood by recentering a static seed solution on an arbitrary timelike geodesic-it is natural to expect that the same strategy should apply to the gravitational field of a massive particle. At the linearized level, the natural starting point is the Schwarzschild--AdS seed at the center of global AdS, which should then be rewritten in coordinates adapted to a general source geodesic, in direct analogy with the electromagnetic Coulomb solution studied here. A particularly promising feature of the gravitational problem is that it can be formulated in terms of gauge-invariant quantities such as the electric part of the Weyl tensor, rather than metric perturbations themselves, thereby providing a closer analogue of the role played by the field strength in the gauge-theory case. Preliminary considerations \cite{Duary:gravitational_extension} suggest that the corresponding future and past null-fringe data again organize themselves into anisotropic Coulombic profiles related by an antipodal map, hinting at an AdS precursor of the familiar gravitational matching conditions in flat-space. It would be especially interesting to understand how these null-fringe Weyl data are encoded holographically, perhaps through an appropriate boundary stress-tensor description, and how their large-\(L\) limit reproduces the asymptotic gravitational data at \(\mathscr I^\pm\). We leave a systematic development of this gravitational extension to future work.

\paragraph{Embedding-space formulation and antipodal matching.}
Another intriguing direction is to develop a fully embedding-space formulation of the problem. The present work already makes essential use of embedding-space invariants in reconstructing the adapted coordinates, but it would be desirable to formulate the Li\'enard--Wiechert solution directly in the Dirac ambient-space language \cite{Dirac:1936fq}, with the gauge field and field strength represented in a manifestly covariant way in $\mathbb{R}^{3,2}$. Such a formulation could provide a considerably more compact description of the solution and might make the origin of antipodal matching even more transparent, potentially as a consequence of a simple discrete transformation in the ambient space. 


\paragraph{Light-ray operators and null-quantized boundary data.}
A natural future direction is to clarify the boundary meaning of the AdS Li\'enard--Wiechert field in terms of light-ray operators. In the present construction, the Coulombic data of a charge moving along a timelike geodesic are transported to the AdS boundary along null rays and become localized near the two distinguished null-fringe regions $\tau=\pm \pi/2$. This suggests that the boundary dual of the leading Coulombic field should not be understood simply as an ordinary local operator at a fixed boundary time, but rather as a light-ray current operator supported on, or sharply localized near, these null-fringe strips. In this language, the bulk component $F_{\rho\tau}$ would be related to an integrated boundary current operator smeared along the future or past null fringe, and the antipodal matching condition would become a relation between the corresponding future and past light-ray operators at antipodal angles. Such a reformulation would provide a purely boundary interpretation of the classical Li\'enard--Wiechert matching condition and would connect the finite-radius AdS construction directly to Carrollian and celestial descriptions of infrared data \cite{Duary:2022onm, Duary:2024fii, Duary:2024cqb}. It would be interesting direction to reformulate the solution using the null-quantized and shadow-completed bases recently developed for Lorentzian AdS in \cite{Navarro:2025xln}. The AdS Li\'enard--Wiechert field is a particularly useful test case because it is an exact classical solution, has a simple geodesic-centered Coulombic origin, and nevertheless carries nontrivial null-fringe data in the flat-space limit. One may ask whether this field admits a natural expansion in AdS boost eigenfunctions \cite{Navarro:2025xln}, whose flat-space limits become conformal-primary wavefunctions, and whether its future and past boundary data are related by a Lorentzian shadow transform. This would also provide a route toward defining finite-radius AdS precursors of soft photon charges, with the flat-space large-gauge Ward identity emerging from an AdS null-fringe conservation law.

\paragraph{Classical Coulombic data as a probe of Lorentzian AdS reconstruction.}
The recent analysis of \cite{Navarro:2026rna} clarifies how Lorentzian AdS
bulk reconstruction already contains the seeds of flat-space celestial and
Carrollian holography. In particular, time-ordered bulk-to-boundary
propagators furnish bases for positive- and negative-energy AdS solutions, and
their large-radius limits reproduce either plane waves or Carrollian
wavefunctions, depending on whether the conformal dimension is scaled with the
AdS radius or held fixed. From this perspective, the boundary regions in the
past or future of a bulk slice become the natural AdS ancestors of the
in/out regions of flat-space scattering.

Our construction supplies a complementary classical realization of this
mechanism for Maxwell theory. The Li\'enard--Wiechert field considered here nevertheless
exhibits the same geometric organization: the Coulombic data of a geodesic
source are carried to the boundary strips near $\tau=\pm\pi/2$, and these
strips become the flat-space null infinities in the large-$L$ limit. The
antipodal relation between the two strips is therefore naturally interpreted
as the Coulombic, infrared counterpart of the Lorentzian AdS-to-Carrollian
map. Thus the present calculation may be regarded as a spin-one, classical
test case for the broader statement that flat-space in/out observables arise
from finite-radius AdS data with specific Lorentzian support.



\section*{Acknowledgements}
I would like to acknowledge support from the Shuimu Tsinghua Scholar Program of Tsinghua University and the Beijing Natural Science Foundation of China, Grant No. IS25035. I would like to thank the organizers of the Pre-Strings School 2026 and Strings 2026, held at the Shanghai Institute for Mathematics and Interdisciplinary Sciences (SIMIS) and Dongjiao State Guest Hotel, Shanghai, for their warm hospitality and for providing vibrant and stimulating academic environments. The final part of this work was completed during these programs, while staying at the WeiTing Hotel and the Dongjiao State Guest Hotel.



\appendix
\section{Image charge at spatial infinity}
\label{app:image-charge-spatial-infinity}
\label{imch}

We now make more precise the statement that, after conformal compactification, a
Coulomb field sourced by an isolated charge in Minkowski space is represented on
the compactified spatial slice by an image singularity at spatial infinity.  Let
\(\widetilde M\subset \mathbb R_T\times S^3\) denote the conformal compactification
of Minkowski space into the Einstein static universe.  A spatial Cauchy slice of
Minkowski space is \(\mathbb R^3\), while its one-point compactification is
\begin{equation}
\label{eq:R3_one_point_compactification}
\mathbb R^3\cup \{i^0\}\simeq S^3 .
\end{equation}
Thus, after compactification, spatial infinity is represented by a single point
\(i^0\in S^3\).  We denote the compactified spatial slice by
\(\Sigma\simeq S^3\), equipped with metric \(h_{ab}\), volume form
\(\mathrm{vol}_{\Sigma}\), Hodge star \(*_{\Sigma}\), and Laplacian
\(\Delta_{\Sigma}=h^{ab}\nabla_a\nabla_b\).  We normalize the delta distribution by
\begin{equation}
\label{eq:delta_normalization}
\int_{\Sigma} f(x)\,\delta_{\Sigma}^{(3)}(x,p)\,
\mathrm{vol}_{\Sigma}
=
f(p)
\end{equation}
for every smooth test function \(f\).

On the compact slice, Gauss's law takes the distributional form
\begin{equation}
\label{eq:gauss_compact_slice_rigorous}
d\,{*}_{\Sigma}E
=
\rho\,\mathrm{vol}_{\Sigma}.
\end{equation}
If \(E\) were smooth on all of \(\Sigma\), then integrating over the compact manifold
would give
\begin{equation}
\label{eq:compact_gauss_zero}
\int_{\Sigma}\rho\,\mathrm{vol}_{\Sigma}
=
\int_{\Sigma}d\,{*}_{\Sigma}E
=
\int_{\partial\Sigma}{*}_{\Sigma}E
=
0,
\end{equation}
because \(\partial\Sigma=\varnothing\).  Therefore a smooth Maxwell field on a
compact spatial slice cannot carry nonzero total electric charge.  This is the
zero-mode obstruction to solving Gauss's law on a compact manifold.

Suppose now that the physical spacetime contains one charge \(q\) at a point
\(p\in \Sigma\setminus\{i^0\}\).  In the original noncompact Minkowski slice
\(\mathbb R^3\), Gauss's law reads
\begin{equation}
\label{eq:gauss_noncompact}
d\,{*}_{\mathbb R^3}E
=
q\,\delta_{\mathbb R^3}^{(3)}(x,p)\,
\mathrm{vol}_{\mathbb R^3}.
\end{equation}
The flux through a large sphere \(S^2_R\subset \mathbb R^3\) is
\begin{equation}
\label{eq:large_sphere_flux}
\lim_{R\to\infty}
\int_{S^2_R}{*}_{\mathbb R^3}E
=
q .
\end{equation}
After compactification, the large sphere \(S^2_R\) shrinks around the added point
\(i^0\).  Thus the flux which escaped through infinity in the noncompact
description becomes, in the compact description, a singular contribution located at
\(i^0\).  Equivalently, Gauss's law on \(\Sigma\simeq S^3\) must be written as
\begin{equation}
\label{eq:image_charge_gauss_law_rigorous}
d\,{*}_{\Sigma}E
=
q\left[
\delta_{\Sigma}^{(3)}(x,p)
-
\delta_{\Sigma}^{(3)}(x,i^0)
\right]\mathrm{vol}_{\Sigma}.
\end{equation}
The second term is the image charge at spatial infinity.  It is not an additional
physical charge in Minkowski space.  Rather, it is the distributional singularity
required to represent a nonzero Coulomb flux on the compactified slice.

The necessity of the image term follows immediately by integrating
\eqref{eq:image_charge_gauss_law_rigorous} over \(\Sigma\)
\begin{equation}
\label{eq:image_charge_total_zero}
\int_{\Sigma}
q\left[
\delta_{\Sigma}^{(3)}(x,p)
-
\delta_{\Sigma}^{(3)}(x,i^0)
\right]\mathrm{vol}_{\Sigma}
=
q-q
=
0.
\end{equation}
Thus the compactified charge distribution has zero total charge, as required on a
compact manifold.

Equivalently, if the electric field is locally written as
\begin{equation}
\label{eq:E_from_potential}
E=-d_{\Sigma}\Phi,
\end{equation}
then, with the sign convention
\(-\Delta_{\Sigma}\Phi\,\mathrm{vol}_{\Sigma}=d{*}_{\Sigma}E\), the potential obeys
the distributional Poisson equation
\begin{equation}
\label{eq:image_charge_poisson_rigorous}
-\Delta_{\Sigma}\Phi
=
q\left[
\delta_{\Sigma}^{(3)}(x,p)
-
\delta_{\Sigma}^{(3)}(x,i^0)
\right].
\end{equation}
The right-hand side has vanishing integral,
\begin{equation}
\label{eq:poisson_zero_mode_condition}
\int_{\Sigma}
q\left[
\delta_{\Sigma}^{(3)}(x,p)
-
\delta_{\Sigma}^{(3)}(x,i^0)
\right]\mathrm{vol}_{\Sigma}
=
0,
\end{equation}
which is precisely the solvability condition for the Laplacian on a compact
manifold.  Therefore the potential can be written in terms of the bipolar Green
function \(G_{\Sigma}(x;p,i^0)\), defined by
\begin{equation}
\label{eq:bipolar_green_definition}
-\Delta_{\Sigma}G_{\Sigma}(x;p,i^0)
=
\delta_{\Sigma}^{(3)}(x,p)
-
\delta_{\Sigma}^{(3)}(x,i^0),
\qquad
\int_{\Sigma}G_{\Sigma}(x;p,i^0)\,\mathrm{vol}_{\Sigma}=0.
\end{equation}
Then
\begin{equation}
\label{eq:potential_bipolar_green}
\Phi(x)
=
q\,G_{\Sigma}(x;p,i^0)
\end{equation}
solves \eqref{eq:image_charge_poisson_rigorous}.  The Green function has the local
Coulomb singularities
\begin{equation}
\label{eq:bipolar_green_local_p}
G_{\Sigma}(x;p,i^0)
=
\frac{1}{4\pi\,d_{\Sigma}(x,p)}
+\mathscr{O}(1),
\qquad x\to p,
\end{equation}
and
\begin{equation}
\label{eq:bipolar_green_local_i0}
G_{\Sigma}(x;p,i^0)
=
-\frac{1}{4\pi\,d_{\Sigma}(x,i^0)}
+\mathscr{O}(1),
\qquad x\to i^0,
\end{equation}
where \(d_{\Sigma}\) is the geodesic distance on \(\Sigma\).  Consequently,
\begin{equation}
\label{eq:E_i0_divergence_rigorous}
|E|
=
|d_{\Sigma}\Phi|
\sim
\frac{|q|}{4\pi\,d_{\Sigma}(x,i^0)^2},
\qquad x\to i^0.
\end{equation}
Thus the compactified Coulomb field is generically singular at spatial infinity.

For intuition, in a conformal frame where the two singularities are placed at the
north and south poles of a unit \(S^3\), one may use the polar angle
\(\chi\in(0,\pi)\).  The bipolar Green function is then represented locally by
\begin{equation}
\label{eq:explicit_bipolar_green_s3}
G_{N,S}(\chi)
=
\frac{1}{4\pi}\cot\chi ,
\end{equation}
which satisfies, in the distributional sense,
\begin{equation}
\label{eq:explicit_bipolar_green_distribution}
-\Delta_{S^3}G_{N,S}
=
\delta^{(3)}_{S^3}(x,N)
-
\delta^{(3)}_{S^3}(x,S).
\end{equation}
Indeed, near \(\chi=0\),
\begin{equation}
\label{eq:green_north_expansion}
G_{N,S}(\chi)
\sim
\frac{1}{4\pi\chi},
\end{equation}
while near the opposite pole, \(\chi=\pi\),
\begin{equation}
\label{eq:green_south_expansion}
G_{N,S}(\chi)
\sim
-\frac{1}{4\pi(\pi-\chi)}.
\end{equation}
This explicitly exhibits the physical charge and the compensating image charge.

The important conclusion is that smoothness at \(i^0\) is too strong a requirement.
The compactified Maxwell field is allowed, and generically expected, to have a
distributional singularity at spatial infinity.  What remains well-defined is not the
pointwise value of the field at \(i^0\), but the limiting Coulombic data transported
along the null generators of conformal infinity.  This is precisely the content of
the antipodal matching condition
\begin{equation}
\label{eq:image-charge-antipodal-matching-condition}
\left.
\lim_{r\to\infty} r^{2}F_{ru}(u,\hat x)
\right|_{\mathscr I^{+}_{-}}
=
\left.
\lim_{r\to\infty} r^{2}F_{rv}(v,-\hat x)
\right|_{\mathscr I^{-}_{+}} .
\end{equation}
In other words, the image-charge singularity explains why the field need not be
smooth at \(i^0\), while the antipodal matching condition identifies the part of the
field that is nevertheless continuous through \(i^0\): the leading Coulombic data
along each null generator of \(\mathscr I\).

\bibliographystyle{JHEP}
\bibliography{ref}

@article{Hijano:2020szl,
    author = "Hijano, Eliot and Neuenfeld, Dominik",
    title = "{Soft photon theorems from CFT Ward identites in the flat limit of AdS/CFT}",
    eprint = "2005.03667",
    archivePrefix = "arXiv",
    primaryClass = "hep-th",
    doi = "10.1007/JHEP11(2020)009",
    journal = "JHEP",
    volume = "11",
    pages = "009",
    year = "2020"
}

@article{Lienard1898,
    author  = {Li{\'e}nard, A.},
    title   = {{Champ {\'e}lectrique et magn{\'e}tique produit par une charge concentr{\'e}e en un point et anim{\'e}e d'un mouvement quelconque}},
    journal = {{L'{\'E}clairage {\'E}lectrique}},
    volume  = {16},
    number  = {27},
    pages   = {5--14},
    year    = {1898}
}

@article{Wiechert1901,
    author  = {Wiechert, E.},
    title   = {{Elektrodynamische Elementargesetze}},
    journal = {Annalen der Physik},
    volume  = {309},
    pages   = {667--689},
    year    = {1901}
}

@book{Schwinger2018,
    author    = {Schwinger, J. and DeRaad, L. L. Jr. and Milton, K. A. and Tsai, W.},
    title     = {{Classical Electrodynamics}},
    publisher = {CRC Press, Taylor \& Francis Group},
    year      = {2018}
}

@book{Panofsky1975,
    author    = {Panofsky, W. K. H. and Phillips, M.},
    title     = {{Classical Electricity and Magnetism}},
    edition   = {2},
    publisher = {Addison-Wesley},
    address   = {Reading},
    year      = {1975}
}

@book{Jackson1998nia,
    author    = {Jackson, John David},
    title     = {{Classical Electrodynamics}},
    edition   = {3},
    isbn      = {978-0-471-30932-1},
    publisher = {Wiley},
    address   = {New York},
    year      = {1998}
}

@article{Minkowski1909,
    author  = {Minkowski, H.},
    title   = {{Raum und Zeit}},
    journal = {Jahresbericht der Deutschen Mathematiker-Vereinigung},
    volume  = {18},
    pages   = {75--88},
    year    = {1909}
}

@article{Padmanabhan:2008gr,
    author        = {Padmanabhan, Hari K.},
    title         = {{A Simple derivation of the electromagnetic field of an arbitrarily moving charge}},
    journal       = {American Journal of Physics},
    volume        = {77},
    pages         = {151--155},
    year          = {2009},
    eprint        = {0810.4246},
    archivePrefix = {arXiv},
    primaryClass  = {physics.class-ph}
}

@book{Zangwill2013,
    author    = {Zangwill, Andrew},
    title     = {{Modern Electrodynamics}},
    publisher = {Cambridge University Press},
    address   = {New York},
    year      = {2013}
}

@incollection{Minkowski1952,
    author    = {Minkowski, H.},
    title     = {{Space and Time}},
    booktitle = {{The Principle of Relativity}},
    editor    = {Lorentz, H. A. and Einstein, A. and Minkowski, H.},
    publisher = {Dover},
    address   = {New York},
    pages     = {75--91},
    year      = {1952}
}

@book{LandauLifshitz1975,
    author    = {Landau, L. D. and Lifshitz, E. M.},
    title     = {{The Classical Theory of Fields}},
    series    = {{Course of Theoretical Physics}},
    volume    = {2},
    edition   = {4},
    publisher = {Pergamon Press},
    address   = {Oxford},
    year      = {1975}
}

@book{Rohrlich2007,
    author    = {Rohrlich, Fritz},
    title     = {{Classical Charged Particles}},
    edition   = {3},
    publisher = {World Scientific},
    address   = {Singapore},
    year      = {2007},
    isbn      = {978-981-270-004-9}
}

@book{Feynman1964,
    author    = {Feynman, Richard P. and Leighton, Robert B. and Sands, Matthew},
    title     = {{The Feynman Lectures on Physics, Vol. II: Mainly Electromagnetism and Matter}},
    publisher = {Addison-Wesley},
    address   = {Reading, MA},
    year      = {1964}
}

@misc{McDonaldLW2017,
    author = {McDonald, Kirk T.},
    title  = {{Li{\'e}nard--Wiechert Potentials and Fields via Lorentz Transformations}},
    note   = {Joseph Henry Laboratories, Princeton University; originally April 7, 1979, updated April 25, 2017},
    year   = {2017},
    url    = {http://kirkmcd.princeton.edu/examples/EM/lw_potentials.pdf}
}

@misc{Fitzpatrick2006,
    author = {Fitzpatrick, Richard},
    title  = {{Potential due to a Moving Charge}},
    note   = {Lecture notes, University of Texas at Austin},
    year   = {2006},
    url    = {http://farside.ph.utexas.edu/teaching/em/lectures/node124.html}
}

@techreport{Streltsov1993,
    author      = {Strel'tsov, V. N.},
    title       = {{Lienard--Wiechert Potential as a Consequence of Lorentz Transformation of Coulomb Potential}},
    institution = {Joint Institute for Nuclear Research},
    number      = {JINR Report D2-93-437},
    year        = {1993}
}

@book{Griffiths2017,
    author    = {Griffiths, David J.},
    title     = {{Introduction to Electrodynamics}},
    edition   = {4},
    publisher = {Cambridge University Press},
    year      = {2017},
    isbn      = {978-1-108-42041-9}
}

@article{Duary:2024kxl,
    author = "Duary, Sarthak and Upadhyay, Shivam",
    title = "{Flat limit of AdS/CFT from AdS geodesics: Scattering amplitudes and antipodal matching of Li{\'e}nard-Wiechert fields}",
    eprint = "2411.08540",
    archivePrefix = "arXiv",
    primaryClass = "hep-th",
    doi = "10.1016/j.nuclphysb.2025.116927",
    journal = "Nucl. Phys. B",
    volume = "1017",
    pages = "116927",
    year = "2025"
}

@article{Klein:2009is,
    author = "Klein, David and Collas, Peter",
    title = "{Exact Fermi coordinates for a class of space-times}",
    eprint = "0912.2779",
    archivePrefix = "arXiv",
    primaryClass = "math-ph",
    doi = "10.1063/1.3298684",
    journal = "J. Math. Phys.",
    volume = "51",
    pages = "022501",
    year = "2010"
}

@article{Allen:1985wd,
    author = "Allen, Bruce and Jacobson, Theodore",
    title = "{Vector two-point functions in maximally symmetric spaces}",
    doi = "10.1007/BF01211169",
    journal = "Commun. Math. Phys.",
    volume = "103",
    pages = "669--692",
    year = "1986"
}

@article{Chu:2019ben,
    author = "Chu, Chong-Sun and Koyama, Yoji",
    title = "{Memory effect in anti--de Sitter spacetime}",
    eprint = "1906.09361",
    archivePrefix = "arXiv",
    primaryClass = "gr-qc",
    doi = "10.1103/PhysRevD.100.104034",
    journal = "Phys. Rev. D",
    volume = "100",
    pages = "104034",
    year = "2019"
}

@book{Strominger:2017zoo,
    author = "Strominger, Andrew",
    title = "{Lectures on the Infrared Structure of Gravity and Gauge Theory}",
    eprint = "1703.05448",
    archivePrefix = "arXiv",
    primaryClass = "hep-th",
    isbn = "978-0-691-17973-5",
    publisher = "Princeton University Press",
    year = "2018"
}

@article{Polchinski:1999ry,
    author = "Polchinski, Joseph",
    title = "{S matrices from AdS space-time}",
    eprint = "hep-th/9901076",
    archivePrefix = "arXiv",
    reportNumber = "NST-ITP-99-02",
    month = "1",
    year = "1999"
}

@article{Susskind:1998vk,
    author = "Susskind, Leonard",
    editor = "Burgess, C. P. and Myers, Robert C.",
    title = "{Holography in the flat space limit}",
    eprint = "hep-th/9901079",
    archivePrefix = "arXiv",
    doi = "10.1063/1.1301570",
    journal = "AIP Conf. Proc.",
    volume = "493",
    number = "1",
    pages = "98--112",
    year = "1999"
}

@article{Giddings:1999qu,
    author = "Giddings, Steven B.",
    title = "{The Boundary S matrix and the AdS to CFT dictionary}",
    eprint = "hep-th/9903048",
    archivePrefix = "arXiv",
    doi = "10.1103/PhysRevLett.83.2707",
    journal = "Phys. Rev. Lett.",
    volume = "83",
    pages = "2707--2710",
    year = "1999"
}

@article{Giddings:1999jq,
    author = "Giddings, Steven B.",
    title = "{Flat space scattering and bulk locality in the AdS / CFT correspondence}",
    eprint = "hep-th/9907129",
    archivePrefix = "arXiv",
    doi = "10.1103/PhysRevD.61.106008",
    journal = "Phys. Rev. D",
    volume = "61",
    pages = "106008",
    year = "2000"
}

@article{Okuda:2010ym,
    author = "Okuda, Takuya and Penedones, Joao",
    title = "{String scattering in flat space and a scaling limit of Yang-Mills correlators}",
    eprint = "1002.2641",
    archivePrefix = "arXiv",
    primaryClass = "hep-th",
    doi = "10.1103/PhysRevD.83.086001",
    journal = "Phys. Rev. D",
    volume = "83",
    pages = "086001",
    year = "2011"
}

@article{Penedones:2010ue,
    author = "Penedones, Joao",
    title = "{Writing CFT correlation functions as AdS scattering amplitudes}",
    eprint = "1011.1485",
    archivePrefix = "arXiv",
    primaryClass = "hep-th",
    doi = "10.1007/JHEP03(2011)025",
    journal = "JHEP",
    volume = "03",
    pages = "025",
    year = "2011"
}

@article{Fitzpatrick:2011jn,
    author = "Fitzpatrick, A. Liam and Kaplan, Jared",
    title = "{Scattering States in AdS/CFT}",
    eprint = "1104.2597",
    archivePrefix = "arXiv",
    primaryClass = "hep-th",
    reportNumber = "SLAC-PUB-14507",
    month = "4",
    year = "2011"
}

@article{Maldacena:2015iua,
    author = "Maldacena, Juan and Simmons-Duffin, David and Zhiboedov, Alexander",
    title = "{Looking for a bulk point}",
    eprint = "1509.03612",
    archivePrefix = "arXiv",
    primaryClass = "hep-th",
    doi = "10.1007/JHEP01(2017)013",
    journal = "JHEP",
    volume = "01",
    pages = "013",
    year = "2017"
}

@article{Hijano:2019qmi,
    author = "Hijano, Eliot",
    title = "{Flat space physics from AdS/CFT}",
    eprint = "1905.02729",
    archivePrefix = "arXiv",
    primaryClass = "hep-th",
    doi = "10.1007/JHEP07(2019)132",
    journal = "JHEP",
    volume = "07",
    pages = "132",
    year = "2019"
}

@article{Li:2021snj,
    author = "Li, Yue-Zhou",
    title = "{Notes on flat-space limit of AdS/CFT}",
    eprint = "2106.04606",
    archivePrefix = "arXiv",
    primaryClass = "hep-th",
    doi = "10.1007/JHEP09(2021)027",
    journal = "JHEP",
    volume = "09",
    pages = "027",
    year = "2021"
}

@article{Duary:2023gqg,
    author = "Duary, Sarthak",
    title = "{Flat limit of massless scalar scattering in AdS2}",
    eprint = "2305.20037",
    archivePrefix = "arXiv",
    primaryClass = "hep-th",
    doi = "10.1016/j.nuclphysb.2024.116687",
    journal = "Nucl. Phys. B",
    volume = "1007",
    pages = "116687",
    year = "2024"
}

@article{Fontanella:2025tbs,
    author = "Fontanella, Andrea and Payne, Oliver",
    title = "{A Carroll Limit of AdS/CFT: A Triality with Flat Space Holography?}",
    eprint = "2508.10085",
    archivePrefix = "arXiv",
    primaryClass = "hep-th",
    month = "8",
    year = "2025"
}

@article{Bekaert:2026cib,
    author = "Bekaert, Xavier and Campoleoni, Andrea and Pekar, Simon and Raj, S. I. Aadharsh",
    title = "{Flat from AdS: in any dimension and for any spin}",
    eprint = "2606.03955",
    archivePrefix = "arXiv",
    primaryClass = "hep-th",
    month = "6",
    year = "2026"
}

@article{Paulos:2016fap,
    author = "Paulos, Miguel F. and Penedones, Joao and Toledo, Jonathan and van Rees, Balt C. and Vieira, Pedro",
    title = "{The S-matrix bootstrap. Part I: QFT in AdS}",
    eprint = "1607.06109",
    archivePrefix = "arXiv",
    primaryClass = "hep-th",
    reportNumber = "CERN-TH-2016-162",
    doi = "10.1007/JHEP11(2017)133",
    journal = "JHEP",
    volume = "11",
    pages = "133",
    year = "2017"
}

@article{Dubovsky:2017cnj,
    author = "Dubovsky, Sergei and Gorbenko, Victor and Mirbabayi, Mehrdad",
    title = "{Asymptotic fragility, near AdS$_{2}$ holography and $ T\overline{T} $}",
    eprint = "1706.06604",
    archivePrefix = "arXiv",
    primaryClass = "hep-th",
    doi = "10.1007/JHEP09(2017)136",
    journal = "JHEP",
    volume = "09",
    pages = "136",
    year = "2017"
}

@article{Carmi:2018qzm,
    author = "Carmi, Dean and Di Pietro, Lorenzo and Komatsu, Shota",
    title = "{A Study of Quantum Field Theories in AdS at Finite Coupling}",
    eprint = "1810.04185",
    archivePrefix = "arXiv",
    primaryClass = "hep-th",
    doi = "10.1007/JHEP01(2019)200",
    journal = "JHEP",
    volume = "01",
    pages = "200",
    year = "2019"
}

@article{Komatsu:2020sag,
    author = "Komatsu, Shota and Paulos, Miguel F. and Van Rees, Balt C. and Zhao, Xiang",
    title = "{Landau diagrams in AdS and S-matrices from conformal correlators}",
    eprint = "2007.13745",
    archivePrefix = "arXiv",
    primaryClass = "hep-th",
    reportNumber = "CPHT-RR119.122020",
    doi = "10.1007/JHEP11(2020)046",
    journal = "JHEP",
    volume = "11",
    pages = "046",
    year = "2020"
}

@article{Antunes:2021abs,
    author = "Antunes, Ant{\'o}nio and Costa, Miguel S. and Penedones, Jo{\~a}o and Salgarkar, Aaditya and van Rees, Balt C.",
    title = "{Towards bootstrapping RG flows: sine-Gordon in AdS}",
    eprint = "2109.13261",
    archivePrefix = "arXiv",
    primaryClass = "hep-th",
    doi = "10.1007/JHEP12(2021)094",
    journal = "JHEP",
    volume = "12",
    pages = "094",
    year = "2021"
}

@article{Gadde:2022ghy,
    author = "Gadde, Abhijit and Sharma, Trakshu",
    title = "{A scattering amplitude for massive particles in AdS}",
    eprint = "2204.06462",
    archivePrefix = "arXiv",
    primaryClass = "hep-th",
    doi = "10.1007/JHEP09(2022)157",
    journal = "JHEP",
    volume = "09",
    pages = "157",
    year = "2022"
}

@article{Cordova:2022pbl,
    author = "C{\'o}rdova, Luc{\'\i}a and He, Yifei and Paulos, Miguel F.",
    title = "{From conformal correlators to analytic S-matrices: CFT$_{1}$/QFT$_{2}$}",
    eprint = "2203.10840",
    archivePrefix = "arXiv",
    primaryClass = "hep-th",
    doi = "10.1007/JHEP08(2022)186",
    journal = "JHEP",
    volume = "08",
    pages = "186",
    year = "2022"
}

@article{Ankur:2023lum,
    author = "Ankur and Carmi, Dean and Di Pietro, Lorenzo",
    title = "{Scalar QED in AdS}",
    eprint = "2306.05551",
    archivePrefix = "arXiv",
    primaryClass = "hep-th",
    doi = "10.1007/JHEP10(2023)089",
    journal = "JHEP",
    volume = "10",
    pages = "089",
    year = "2023"
}

@article{Duary:2022pyv,
    author = "Duary, Sarthak and Hijano, Eliot and Patra, Milan",
    title = "{Towards an IR finite S-matrix in the flat limit of AdS/CFT}",
    eprint = "2211.13711",
    archivePrefix = "arXiv",
    primaryClass = "hep-th",
    month = "11",
    year = "2022"
}

@article{Banerjee:2022oll,
    author = "Banerjee, Nabamita and Fernandes, Karan and Mitra, Arpita",
    title = "{1/L$^{2}$ corrected soft photon theorem from a CFT$_{3}$ Ward identity}",
    eprint = "2209.06802",
    archivePrefix = "arXiv",
    primaryClass = "hep-th",
    doi = "10.1007/JHEP04(2023)055",
    journal = "JHEP",
    volume = "04",
    pages = "055",
    year = "2023"
}

@article{Duary:2022afn,
    author = "Duary, Sarthak",
    title = "{AdS correction to the Faddeev-Kulish state: migrating from the flat peninsula}",
    eprint = "2212.09509",
    archivePrefix = "arXiv",
    primaryClass = "hep-th",
    doi = "10.1007/JHEP05(2023)079",
    journal = "JHEP",
    volume = "05",
    pages = "079",
    year = "2023"
}

@article{Banerjee:2024yiq,
    author = "Banerjee, Nabamita and Desai, Amogh Neelkanth and Fernandes, Karan and Mitra, Arpita and Rahnuma, Tabasum",
    title = "{AdS S-matrix for massive vector fields}",
    eprint = "2412.19253",
    archivePrefix = "arXiv",
    primaryClass = "hep-th",
    doi = "10.1007/JHEP05(2025)094",
    journal = "JHEP",
    volume = "05",
    pages = "094",
    year = "2025"
}

@article{deGioia:2022fcn,
    author = "de Gioia, Leonardo Pipolo and Raclariu, Ana-Maria",
    title = "{Eikonal approximation in celestial CFT}",
    eprint = "2206.10547",
    archivePrefix = "arXiv",
    primaryClass = "hep-th",
    doi = "10.1007/JHEP03(2023)030",
    journal = "JHEP",
    volume = "03",
    pages = "030",
    year = "2023"
}

@article{Bagchi:2023fbj,
    author = "Bagchi, Arjun and Dhivakar, Prateksh and Dutta, Sudipta",
    title = "{AdS Witten diagrams to Carrollian correlators}",
    eprint = "2303.07388",
    archivePrefix = "arXiv",
    primaryClass = "hep-th",
    doi = "10.1007/JHEP04(2023)135",
    journal = "JHEP",
    volume = "04",
    pages = "135",
    year = "2023"
}

@article{deGioia:2024yne,
    author = "de Gioia, Leonardo Pipolo and Raclariu, Ana-Maria",
    title = "{Celestial amplitudes from conformal correlators with bulk-point kinematics}",
    eprint = "2405.07972",
    archivePrefix = "arXiv",
    primaryClass = "hep-th",
    month = "5",
    year = "2024"
}

@article{Alday:2024yyj,
    author = "Alday, Luis F. and Nocchi, Maria and Ruzziconi, Romain and Yelleshpur Srikant, Akshay",
    title = "{Carrollian amplitudes from holographic correlators}",
    eprint = "2406.19343",
    archivePrefix = "arXiv",
    primaryClass = "hep-th",
    doi = "10.1007/JHEP03(2025)158",
    journal = "JHEP",
    volume = "03",
    pages = "158",
    year = "2025"
}

@article{Li:2024kbo,
    author = "Li, Ang and Long, Jiang and Yang, Jing-Long",
    title = "{Carrollian propagator and amplitude in Rindler spacetime}",
    eprint = "2410.20372",
    archivePrefix = "arXiv",
    primaryClass = "hep-th",
    doi = "10.1007/JHEP03(2025)186",
    journal = "JHEP",
    volume = "03",
    pages = "186",
    year = "2025"
}

@article{Navarro:2025xln,
    author = "Navarro, N{\'u}ria and Raclariu, Ana-Maria",
    title = "{Null quantization, shadows and boost eigenfunctions in Lorentzian AdS}",
    eprint = "2512.13541",
    archivePrefix = "arXiv",
    primaryClass = "hep-th",
    month = "12",
    year = "2025"
}

@article{Bagchi:2026emg,
    author = "Bagchi, Arjun and Lipstein, Arthur and Mondal, Saikat and Zhang, Alex Jiayi",
    title = "{Carrollian ABJM: Fermions and Supersymmetry}",
    eprint = "2604.22582",
    archivePrefix = "arXiv",
    primaryClass = "hep-th",
    month = "4",
    year = "2026"
}

@article{Dirac:1936fq,
    author = "Dirac, Paul A. M.",
    title = "{Wave equations in conformal space}",
    doi = "10.2307/1968455",
    journal = "Annals Math.",
    volume = "37",
    pages = "429--442",
    year = "1936"
}

@unpublished{Duary:gravitational_extension,
    author = "Duary, Sarthak",
    title = "{Geodesic-centered construction to gravitational data in AdS}",
    note = "Work in progress",
    year = "\textit{to appear 2026}"
}

@article{Berenstein:2025tts,
    author = "Berenstein, David and Simon, Joan",
    title = "{Aspects of the bulk flat space limit in AdS/CFT}",
    eprint = "2510.23697",
    archivePrefix = "arXiv",
    primaryClass = "hep-th",
    month = "10",
    year = "2025"
}

@article{Berenstein:2025qhb,
    author = "Berenstein, David and Li, Ziyi",
    title = "{Spinning Fields in Lorentzian AdS}",
    eprint = "2511.15780",
    archivePrefix = "arXiv",
    primaryClass = "hep-th",
    month = "11",
    year = "2025"
}

@article{Adamo:2025bfr,
    author = "Adamo, Tim and Surubaru, Iustin and Zhu, Bin",
    title = "{From AdS correlators to Carrollian amplitudes with the scattering equations}",
    eprint = "2512.03677",
    archivePrefix = "arXiv",
    primaryClass = "hep-th",
    doi = "10.1007/JHEP02(2026)198",
    journal = "JHEP",
    volume = "02",
    pages = "198",
    year = "2026"
}

@article{Lam:2017ofc,
    author = "Lam, Ho Tat and Shao, Shu-Heng",
    title = "{Conformal Basis, Optical Theorem, and the Bulk Point Singularity}",
    eprint = "1711.06138",
    archivePrefix = "arXiv",
    primaryClass = "hep-th",
    doi = "10.1103/PhysRevD.98.025020",
    journal = "Phys. Rev. D",
    volume = "98",
    number = "2",
    pages = "025020",
    year = "2018"
}

@article{Compere:2019bua,
    author = "Comp{\`e}re, Geoffrey and Fiorucci, Adrien and Ruzziconi, Romain",
    title = "{The $\Lambda$-BMS$_4$ group of dS$_4$ and new boundary conditions for AdS$_4$}",
    eprint = "1905.00971",
    archivePrefix = "arXiv",
    primaryClass = "gr-qc",
    doi = "10.1088/1361-6382/ab3d4b",
    journal = "Class. Quant. Grav.",
    volume = "36",
    number = "19",
    pages = "195017",
    year = "2019",
    note = "[Erratum: Class.Quant.Grav. 38, 229501 (2021)]"
}

@article{Kapec:2017tkm,
    author = "Kapec, Daniel and Perry, Malcolm and Raclariu, Ana-Maria and Strominger, Andrew",
    title = "{Infrared Divergences in QED, Revisited}",
    eprint = "1705.04311",
    archivePrefix = "arXiv",
    primaryClass = "hep-th",
    doi = "10.1103/PhysRevD.96.085002",
    journal = "Phys. Rev. D",
    volume = "96",
    number = "8",
    pages = "085002",
    year = "2017"
}

@article{Choi:2017ylo,
    author = "Choi, Sangmin and Akhoury, Ratindranath",
    title = "{BMS Supertranslation Symmetry Implies Faddeev-Kulish Amplitudes}",
    eprint = "1712.04551",
    archivePrefix = "arXiv",
    primaryClass = "hep-th",
    doi = "10.1007/JHEP02(2018)171",
    journal = "JHEP",
    volume = "02",
    pages = "171",
    year = "2018"
}

@article{Navarro:2026rna,
    author = "Navarro, N{\'u}ria and Raclariu, Ana-Maria",
    title = "{On bulk reconstruction in Lorentzian AdS and its flat space limit}",
    eprint = "2605.16641",
    archivePrefix = "arXiv",
    primaryClass = "hep-th",
    month = "5",
    year = "2026"
}

@article{Hannesdottir:2019umk,
    author = "Hannesdottir, Holmfridur and Schwartz, Matthew D.",
    title = "{Finite $S$ matrix}",
    eprint = "1906.03271",
    archivePrefix = "arXiv",
    primaryClass = "hep-th",
    doi = "10.1103/PhysRevD.107.L021701",
    journal = "Phys. Rev. D",
    volume = "107",
    number = "2",
    pages = "L021701",
    year = "2023"
}

@article{Hannesdottir:2019opa,
    author = "Hannesdottir, Holmfridur and Schwartz, Matthew D.",
    title = "{$S$ -Matrix for massless particles}",
    eprint = "1911.06821",
    archivePrefix = "arXiv",
    primaryClass = "hep-th",
    doi = "10.1103/PhysRevD.101.105001",
    journal = "Phys. Rev. D",
    volume = "101",
    number = "10",
    pages = "105001",
    year = "2020"
}

@article{Ruzziconi:2026bix,
    author = "Ruzziconi, Romain",
    title = "{Carrollian physics and holography}",
    eprint = "2602.02644",
    archivePrefix = "arXiv",
    primaryClass = "hep-th",
    doi = "10.1016/j.physrep.2026.03.005",
    journal = "Phys. Rept.",
    volume = "1182",
    pages = "1--87",
    year = "2026"
}

@article{Zhu:2026ofh,
    author = "Zhu, Bin",
    title = "{Topics in Celestial holography: A bottom-up perspective}",
    eprint = "2606.24285",
    archivePrefix = "arXiv",
    primaryClass = "hep-th",
    month = "6",
    year = "2026"
}

@article{Duary:2022onm,
    author = "Duary, Sarthak",
    title = "{Celestial amplitude for 2d theory}",
    eprint = "2209.02776",
    archivePrefix = "arXiv",
    primaryClass = "hep-th",
    doi = "10.1007/JHEP12(2022)060",
    journal = "JHEP",
    volume = "12",
    pages = "060",
    year = "2022"
}

@phdthesis{Duary:2024fii,
    author = "Duary, Sarthak",
    title = "{Aspects of celestial amplitude and flat-space limit of AdS/CFT}",
    school = "ICTS, Bangalore",
    year = "2024"
}

@article{Duary:2024cqb,
    author = "Duary, Sarthak and Maji, Sourav",
    title = "{Spectral representation in Klein space: simplifying celestial leaf amplitudes}",
    eprint = "2406.02342",
    archivePrefix = "arXiv",
    primaryClass = "hep-th",
    doi = "10.1007/JHEP08(2024)079",
    journal = "JHEP",
    volume = "08",
    pages = "079",
    year = "2024"
}

@article{Diaz:2026igh,
    author = {Diaz, Felipe and H{\"u}sn{\"u}gil, Sercan and Labrin, Oriana and Sanhueza, Leonardo},
    title = "{Gravitational Memory Beyond Null Infinity through Finite-Distance Carrollian Screens}",
    eprint = "2607.18675",
    archivePrefix = "arXiv",
    primaryClass = "hep-th",
    month = "7",
    year = "2026"
}

@article{Compere:2025tzr,
    author = "Comp{\`e}re, Geoffrey and Fontaine, Dima and Nguyen, Kevin",
    title = "{Electromagnetic multipole expansions and the logarithmic soft photon theorem}",
    eprint = "2503.23937",
    archivePrefix = "arXiv",
    primaryClass = "hep-th",
    doi = "10.21468/SciPostPhysCore.8.4.066",
    journal = "SciPost Phys. Core",
    volume = "8",
    pages = "066",
    year = "2025"
}

@article{Fernandes:2023xim,
    author = "Fernandes, Karan and Banerjee, Nabamita and Mitra, Arpita",
    title = "{Soft factors with AdS radius corrections}",
    eprint = "2310.19299",
    archivePrefix = "arXiv",
    primaryClass = "hep-th",
    doi = "10.22128/jhap.2023.738.1059",
    journal = "JHAP",
    volume = "3",
    number = "4",
    pages = "5--22",
    year = "2023"
}

\end{document}